\documentclass[12pt]{article}
\pdfoutput=1
\usepackage{latexsym}
\usepackage{amsmath}
\usepackage{amssymb}
\usepackage[usenames]{color}
\usepackage{tikz}
\usetikzlibrary{decorations.markings}
\usetikzlibrary{snakes} 
\newcommand{\Cross}{$\mathbin{\tikz [x=1.4ex,y=1.4ex,line width=.2ex, red] \draw (0,0) -- (0.8,0.8) (0,0.8) -- (0.8,0);}$}
\definecolor{darkgreen}{rgb}{0,0.5,0}
\definecolor{darkblue}{cmyk}{0.9,0.9,0,0}
\definecolor{darkred}{rgb}{0.6,0,0.3}
\usepackage{cite}
\def\sl#1{\langle #1  \rangle}
\def\del{\partial}
\def\delbar{\bar{\partial}}

\renewcommand{\Im}{{\rm Im}}
\renewcommand{\Re}{{\rm Re}}
\def\barz{\bar{z}}
\def\fn#1{\footnote{#1}}
\def\nn{\nonumber}
\def\eqref#1{(\ref{#1})}
\def\comma{\,,}
\def\period{\,.}
\def\figref#1{Figure \ref{#1}}
\def\bfall{\boldmath\bf}
\def\beq#1{\begin{align}#1\end{align}}
\def\pmatrix#1#2{\left( 
\begin{array}{#1}
#2\end{array} 
\right)}
\numberwithin{equation}{section}
\numberwithin{figure}{section}
\numberwithin{table}{section}
\makeatletter
\def\section{\@startsection {section}{1}{\z@}{-3.5ex plus -1ex minus 
-.2ex}{2.3ex plus .2ex}{\large\bf}}
\makeatother
\makeatletter
\def\subsection{\@startsection {subsection}{1}{\z@}{-3.5ex plus -1ex minus 
-.2ex}{2.3ex plus .2ex}{\normalsize\bf}}
\makeatother
\renewenvironment{thebibliography}{\pagebreak[3]\par\vspace{0.6em}
\begin{flushleft}{\large \bf References}\end{flushleft}
\vspace{-1.0em}

\begin{enumerate}\if@twocolumn\baselineskip=0.6em\itemsep -0.2em
\else\itemsep -0.2em\fi\labelsep 0.1em}{\end{enumerate} }
\usepackage{ifpdf}
\ifpdf
\usepackage{graphicx} 
\usepackage{epsfig}
\usepackage[setpagesize=false,pagebackref=false, linktocpage, bookmarksopen=true, colorlinks=true, linkcolor=darkblue,citecolor=darkblue,urlcolor=darkblue]{hyperref}
\usepackage{hyperref}
\newcommand{\arXiv}[2]{\href{http://arxiv.org/abs/#1}{{\tt arXiv:#2}}}
\newcommand{\hep}[2]{\href{http://arxiv.org/abs/#1}{{\tt #2}}}

\else
\usepackage{graphicx}
\newcommand{\arXiv}[2]{{\tt arXiv:#2}}
\newcommand{\hep}[2]{{\tt #2}}

\fi
\def\Komabanumber#1#2#3{\hfill \begin{minipage}{4.2cm} {\tt UT-Komaba #1}
              \par\noindent {\tt #2} 
              \par\noindent #3 \end{minipage}}
\begin{document}
%%%%%%%%%%%%%%%%%%%%%%%%%%%%

\thispagestyle{empty}

\renewcommand{\thefootnote}{\fnsymbol{footnote}}
\setcounter{page}{1}
\setcounter{footnote}{0}
\setcounter{figure}{0}
\Komabanumber{13-15}{March, 2014}{}
\begin{center}
$$$$
{\Large\textbf{\mathversion{bold}
Classical Liouville Three-point Functions from Riemann-Hilbert Analysis
}\par}

\vspace{1.0cm}

\textrm{Daigo Honda\hyperlink{Honda}{$^{\,\mathcal{X}}$} and Shota Komatsu\hyperlink{Komatsu}{$^{\,\mathcal{Y}}$}}
\\ \vspace{0.6cm}
\footnotesize{
\textit{Institute of Physics, University of Tokyo, \\
 Komaba, Meguro-ku, Tokyo 153-8902, Japan}\\
\vspace{3mm}
{\hypertarget{Honda}{$^{\mathcal{X}}$}} \texttt{dhonda@hep1.c.u-tokyo.ac.jp} \\
{\hypertarget{Komatsu}{$^{\mathcal{Y}}$}}  \texttt{skomatsu@hep1.c.u-tokyo.ac.jp} \\
\vspace{4mm}
}

\par\vspace{1.7cm}

\textbf{\large{Abstract}}\vspace{2mm}
\end{center}
We study semiclassical correlation functions in Liouville field theory on a two-sphere when all operators have large conformal dimensions.
In the usual approach, such computation involves solving the classical Liouville equation, which is known to be extremely difficult for higher-point functions.
To overcome this difficulty, we propose a new method based on the Riemann-Hilbert analysis, which is applied recently to the holographic calculation of correlation functions in AdS/CFT.
The method allows us to directly compute the correlation functions without solving the Liouville equation explicitly.
To demonstrate its utility, we apply it to three-point functions, which are known to be solvable, and confirm that it correctly reproduces the classical limit of the DOZZ formula for quantum three-point functions.
This provides good evidence for the validity of this method.
\noindent
\setcounter{page}{1}
\renewcommand{\thefootnote}{\arabic{footnote}}
\setcounter{footnote}{0}
\newpage
%%%%%%%%%%%%%%%%%%%%%%
\baselineskip 3.5ex
%%%%%%%%table-of-contents%%%%%%%%
\thispagestyle{empty}
\enlargethispage{2\baselineskip}
\tableofcontents
%%%%%%%%%%%%%%%%%%%%%%%%%
\section{Introduction\label{Sec-Int}}

Liouville field theory is one of a few well-studied irrational conformal field theories and has been studied thoroughly in various context such as noncritical string theory, two-dimensional quantum gravity, tachyon condensation and quantum gravity in the de-Sitter spacetime (see comprehensive reviews \cite{GM, Nak} and references therein).
Furthermore, it has been drawing renewed attention recently because of its relation to four-dimensional $\mathcal{N}=2$ supersymmetric gauge theories, known as the AGT correspondence \cite{AGT}.

In this paper, we consider the semiclassical limit of the Liouville field theory, in which the central charge goes to positive infinity. In particular, we discuss semiclassical correlation functions on a two-sphere when all vertex operators have large conformal dimensions. In the AGT correspondence, the semiclassical limit of the Liouville field theory is known to correspond to a special limit of the gauge theory, called the Nekrasov-Shatashvilli limit \cite{NS}. In this limit, an intriguing relation between the gauge theories and the quantum integrable models has been discovered \cite{NS,MiroMoro,MiroMoroShak,NekWit,Tes2010,MaruTaki,Pogh,MarsMiroMoro,NekRosSha,DLH,CDHL,FMPP,BMT,Bour1,BCGK,FMP,Bour2,CHK,MY,NPS}. Although this limit is extensively studied in the literature, the nature and the mechanism of this mysterious relation are still to be elucidated. Thus, for deeper understanding of this relation, it would be worthwhile to investigate the structure of the semiclassical Liouville theory more in detail.

To compute the semiclassical correlation functions in the usual method, we need the explicit form of the solution to the classical Liouville equation. 
However, the solution to the Liouville equation is known only for three-point functions\cite{ZZ,HJ,HMW}, and it is considered to be extremely difficult to obtain the solutions for higher-point functions\fn{Although one can write down the general form of the solutions to the Liouville equation in terms of a meromorphic function and its complex conjugate (see for instance \cite{Seiberg}), it is difficult to determine their explicit form in the case of higher-point functions.}.
To overcome this difficulty, we propose a new method, which does not necessitate the explicit solution to the Liouville equation and is based on a kind of Riemann-Hilbert method developed recently for the holographic calculation of correlation functions in AdS/CFT
\cite{JW, KK1,KK2,CT,KK3}. To demonstrate its utility, we study the three-point functions in this paper and show that it reproduces the known results correctly. 

The key of the method is a certain second-order ordinary differential equation\fn{This differential equation originates from the null-vector decoupling equation in the quantum Liouville field theory and is often referred to as the {\it oper} in the literature \cite{Frenkel}.} associated with the solution to the classical Liouville equation. This equation plays a pivotal role in the analysis since the classical action of the Liouville field theory can be re-expressed in terms of the Wronskians of this differential equation. To compute such Wronskians, we take the following steps. First, from the simple fact that there is no insertion of vertex operators at infinity, we determine certain products of the Wronskians. Next, to compute the individual Wronskians, not just their products, we introduce an extra parameter called spectral parameter and consider a one-parameter deformation of the Wronskians. Then, we discuss the analyticity of the Wronskians with respect to the spectral parameter using a newly introduced quantity called exact WKB curves, which is a generalization of the ordinary WKB curves defined in \cite{GMN2}. With the exact WKB curves, we can fully determine the analyticity of the Wronskians. Then, using the analytic properties thus obtained, we set up the Riemann-Hilbert problem and solve it explicitly in terms of gamma functions. The three-point functions computed in this way turn out to agree completely with the classical limit of the DOZZ three-point functions\cite{ZZ, DO, Tes}. The agreement provides strong evidence for the validity and the utility of this method.

The rest of this paper is organized as follows.
Firstly, in section \ref{Sec-Cl}, we will give a brief summary on the semiclassical limit of the Liouville correlation functions. In particular, we explain that the classical action can be expressed in terms of the Wronskians of certain ordinary differential equations. Secondly, in section \ref{Sec-Wr}, we will determine the Wronskians using the exact WKB curves and the Riemann-Hilbert analysis, and compute the three-point functions exactly. Finally, in section \ref{Sec-Di}, we will conclude and indicate several future directions including the generalization
to the higher-point functions and the relation to the four-dimensional $\mathcal{N}=2$ supersymmetric gauge theories. 
\section{Classical Liouville correlation functions}
\label{Sec-Cl}
%%%%%%
%In this section we add "Cl" to the head of labels
%%%%%%

In this section, we summarize the basics of semiclassical analysis of the Liouville field theory. First, starting from the fully quantum path integral expression of the Liouville field theory, we show that the semiclassical approximation is valid when vertex operators have large conformal dimensions.
Then, we move onto the three-point functions. The semiclassical limit of the three-point functions is a well-studied subject (see \cite{ZZ, HJ,HMW}). However, our approach is different from the previous works as we do not make use of the classical solutions.
After reviewing the long-known fact that the classical solutions of the Liouville equation can be constructed from the solutions of a certain second-order ordinary differential equation, we show that the semiclassical correlation functions can be re-expressed in terms of the Wronskians of this differential equation. Such Wronskians will be directly evaluated in the next section.

\subsection{Semiclassical analysis of Liouville field theory}

The path integral expression of the Liouville correlation function on a two-sphere is given as follows\fn{This action, which is slightly different from the usual one, is obtained after the field redefinition $\phi\to\phi/b$. For a further detail, see, for instance, \cite{HMW}.}
\beq{
\langle\mathcal{V}_1(z_1)\mathcal{V}_2(z_2)\cdots \mathcal{V}_n(z_n) \rangle = \int \mathcal{D}\phi \exp \left[-\frac{1}{2\pi b^2}\int d^2 z \left( \del \phi \delbar \phi + 2\lambda e^{2 \phi}\right)\right]\prod_{i=1}^{n}\mathcal{V}_i(z_i)\comma \label{CL-Quantum}
}
where $\mathcal{V}_i$ is a vertex operator given by
\beq{
\mathcal{V}_i(z_i)=e^{2\alpha_i \phi (z_i)/b}\period
}
Precisely speaking, the boundary term at infinity is necessary to make the action integral convergent \cite{ZZ,HJ,HMW}.
However, in what follows, we will not evaluate the action directly and the precise form of the boundary term will not be needed.

Let us now discuss the semiclassical limit.
The semiclassical limit of the Liouville field theory is basically $b\to 0$ limit.
As \eqref{CL-Quantum} shows, the action part has $1/b^2$ prefactor, which diverges in the limit.
Therefore, the path integral in this limit is dominated by its saddle-point value.
However, if we take the naive $b\to 0$ limit with $\alpha_i$ fixed, the vertex operators in the limit scale as $1/b$, which is $b(\to 0)$ times smaller than the action part.
Then, the saddle point will be insensitive to the vertex operators and the result will become trivial\fn{In such cases, the quantum correction (or equivalently the one-loop correction) will become important.}.
To obtain a nontrivial result, we need to properly scale $\alpha_i$ so that $\eta_i\equiv b \alpha_i$ stays finite.
We call the operators whose conformal dimensions are scaled in this way ``heavy" operators.
In such a limit, the path integral \eqref{CL-Quantum} can be evaluated as
\beq{
&\langle\mathcal{V}_1(z_1)\mathcal{V}_2(z_2)\cdots \mathcal{V}_n(z_n)\rangle 
 \overset{b\to 0}{\sim} \exp \left(-\frac{S}{b^2} \right)\comma \\
&S\equiv \frac{1}{2\pi}\int d^2 z \left( \del \phi_{\ast} \delbar \phi_{\ast} + 2\lambda e^{2\phi_{\ast}} - 4\pi \sum_i \eta _i \phi_{\ast} \delta^2 (z-z_i) \right) \comma
} 
where $\phi_{\ast}$ is the saddle-point value of $\phi$ which satisfies the classical Liouville equation,
\beq{
\del \delbar \phi_{\ast} =\lambda e^{2\phi_{\ast}}-\pi \sum_i \eta _i \delta^2 (z-z_i) \period \label{CL-EOM}
}
In what follows,
we only consider the case where $\eta_i$, which parametrize the vertex operators, are real-valued and smaller than $1/2$.
When $\eta_i$ satisfy these conditions, the exponential term in \eqref{CL-EOM} can be neglected\fn{If we consider the operator with $\eta>1/2$, the asymptotic behavior will be affected by the exponential term in \eqref{CL-EOM}.} in the vicinity of the vertex operators and the asymptotic behavior of $\phi_{\ast}$ is given by
\beq{
\phi_{\ast} \overset{z\to z_i}{\sim} -2 \eta_i \ln |z-z_i| + C_i + O(z-z_i) \period \label{CL-AsympSol}
}
The inequality $\eta_i<1/2$ is known as the {\it Seiberg bound} \cite{Seiberg} in the literature. The modern interpretation\fn{In the context of two-dimensional gravity, the Liouville field plays the role of the metric $g_{ij}=e^{2\phi}\delta_{ij}$ on the two-sphere in the conformal gauge. Then,
the insertion point of a vertex operator with $\eta$ becomes a singularity with a deficit angle $2\pi(1-2\eta)$ and the Seiberg bound $\eta<1/2$ naturally follows from the geometric requirement.} of this bound is that, in the quantum Liouville field theory, the operator with $\eta>1/2$ is equivalent to the operator with $1-\eta$ by some rescaling factor, called the reflection coefficient.

There is another constraint we impose on the values of $\eta_i$.
Integrating \eqref{CL-EOM} over the two-sphere and using the Gauss-Bonnet theorem, we obtain
\beq{
\sum_i \eta_i-1=\frac{\lambda}{\pi} (\textrm{Area}) \comma\label{CL-Area}
}
where $(\textrm{Area})$ denotes the area of the Riemann surface computed with the metric $g_{ij}=e^{\phi_{\ast}}\delta_{ij}$. Therefore, when the Liouville equation has a real-valued solution, \eqref{CL-Area} must be positive.
This provides the necessary condition for the existence of the real solution, which we impose throughout this paper. 
In summary, we require the parameters $\eta_i$ to be in the following region, which we call the {\it physical region},
\beq{
\eta_i < \frac{1}{2} \comma\quad \sum_i \eta_i>1 \period\label{phys-reg}
}
Note that, for the three-point functions, $\eta_i>0$ follows from the above conditions.

Now, using \eqref{CL-EOM}, we can define the following holomorphic quantity\fn{We can also define the anti-holomorphic quantity which corresponds to the anti-chiral stress energy tensor by $\bar{T}(\barz)\equiv -(\delbar \phi_{\ast})^2 +\delbar^2 \phi_{\ast}$.}:
\beq{
T(z)\equiv -(\del \phi_{\ast})^2 +\del^2 \phi_{\ast} \period
}
This quantity is, in fact, a semiclassical limit of the stress-energy tensor and plays an important role in the subsequent analysis.
From \eqref{CL-AsympSol}, the asymptotic behavior of $T(z)$ can be determined in the following form:
\beq{
T(z)=\sum_i \frac{\eta_i(1-\eta_i)}{(z-z_i)^2} + \frac{a_i}{z-z_i} \period
}
Here, the parameters $a_i$ are called accessory parameters. They are constrained by the condition that the stress-energy tensor is not singular and decays as $T(z) \sim z^{-4}$ at infinity in the following way:
\beq{
\sum_i a_i=0,\ \sum_i(a_i z_i + \eta_i(1-\eta_i))=0,\ \sum_i(a_i z_i^2 + 2\eta_i(1-\eta_i)z_i)=0\period\label{accessory}
}
In the case of the three-point functions, \eqref{accessory} is restrictive enough to fully determine the form of $T(z)$ as
\beq{
&T(z)=\left(\frac{\eta_1 (1-\eta_1)z_{12}z_{13}}{z-z_1}+\frac{\eta_2 (1-\eta_2)z_{21}z_{23}}{z-z_2} +\frac{\eta_3 (1-\eta_3)z_{31}z_{32}}{z-z_3}\right)\frac{1}{(z-z_1)(z-z_2)(z-z_3)}\comma
}
where $z_{ij}$ is given by $z_i-z_j$.

Let us now explain how to compute the semiclassical correlation functions. The most straightforward way to compute them is to evaluate the integrand of the path integral \eqref{CL-Quantum}, which consist of the action and the vertex operators, on the saddle-point classical solution. Although this line of approach was taken in the study of correlation functions in AdS/CFT,  there is much an easier way for the Liouville field theory. The idea is to consider the variation of the semiclassical correlation functions with respect to the parameters of the vertex operators, $\eta_i$.  The correlation functions evaluated on the saddle-point depends on $\eta_i$ in two different ways: First, the integrand has an explicit dependence on $\eta_i$, $\exp(2\eta_i\phi /b^2)$. Second, as the saddle-point solution itself depends on $\eta_i$, the integrand depends on $\eta_i$ implicitly through the saddle-point solution. However, the second dependence is always of the form,
\beq{
\frac{\delta S}{\delta \phi} \frac{\delta \phi}{\delta \eta_i} \delta \eta_i\comma \label{CL-EtaVar}
}
which vanishes since the saddle point satisfies the equation of motion.
Therefore, we conclude the only dependence we need to consider is the explicit dependence on $\eta_i$ in the integrand. This way, we arrive at the following important formula:
\beq{
\frac{\del S}{\del \eta_i} =-2 C_i \comma \label{CL-ActionC}
}
where $C_i$ is the subleading term\fn{Although the leading term in \eqref{CL-AsympSol} produces the divergent contribution, it is absorbed by the renormalization of the vertex operators. For details, see \cite{ZZ}.} in \eqref{CL-AsympSol}.
Consequently, the evaluation of the semiclassical limit of the correlation functions boils down to the evaluation of $C_i$.

\subsection{Correlation functions and Wronskians}

In the previous works, $C_i$ is determined by using the explicit solutions of the classical Liouville equation \cite{ZZ,HJ,HMW}.
However, it is extremely difficult to obtain the explicit solutions for the higher-point functions.
To overcome this problem, we will propose another method based on the Riemann-Hilbert analysis in this paper.
The purpose of this section is to express $C_i$ in \eqref{CL-ActionC} in a form to which our method is readily applicable.

First, we use the following famous fact of the classical Liouville field theory: The classical solution of the Liouville field theory can be expressed as the bilinear of the solutions to the ordinary differential equation,
\beq{
\left( \del^2 +T(z)\right) \psi _i =0\comma \quad \left( \delbar^2 +\bar{T}(\barz )\right) \bar{\psi}_i =0\comma \label{CL-AuxLinProb}
}
 as follows:
\beq{
&e^{-\phi} =\sqrt{\lambda}(  \psi_1 \bar{\psi}_1-\psi_2 \bar{\psi}_2) \comma\label{CL-Reconst}\\
&\langle \psi_1 \comma \psi_2 \rangle\langle \bar{\psi}_1 \comma \bar{\psi}_2 \rangle=1\comma
}
where $\langle A\comma B \rangle$ is a Wronskian between $A$ and $B$, defined by
$A \partial B - B \partial A$ or $A \bar{\partial} B - B \bar{\partial} A$.
Existence of such a linear differential equation is one of the important characteristics of integrable systems and plays a pivotal role in the context of the holographic calculation of correlation functions in AdS/CFT \cite{JW, KK1,KK2,CT,KK3}. 
Following the convention in that context, we will call the equation \eqref{CL-AuxLinProb} the {\it auxiliary linear problem}.

Let us now make an important remark on \eqref{CL-Reconst}.
Although it is always true that the classical solution of the Liouville field theory can be represented by a certain bilinear of the solutions to the auxiliary linear problem, not all the bilinears provide a consistent solution of the Liouville field theory.
This is because, while the left hand side of \eqref{CL-Reconst}, $e^{-\phi}$, is always single-valued, the single-valuedness is not guaranteed if we choose a bilinear arbitrarily.
Thus, the correct statement is as follows: {\it If the bilinear of the auxiliary linear problem, defined by the right hand side of \eqref{CL-Reconst}, is single-valued on the Riemann surface we consider, it provides a classical solution of the Liouville field theory.}

Having clarified the relation between the solution of the Liouville field theory and the solutions to the auxiliary linear problem, let us discuss the asymptotic property near the vertex operators.
In the vicinity of the vertex operator, $z_i$, there are two independent solutions $i_{\pm}$ to the auxiliary linear problem \eqref{CL-AuxLinProb} with different asymptotic behavior.
\beq{
&i_{+} \sim (z-z_i)^{\eta_i} \comma \quad i_{-}\sim (z-z_i)^{1-\eta_i}/ (1-2\eta_i) \comma \label{CL-HolAsympt}\\
&\bar{i}_{+}\sim (\barz -\barz_i)^{\eta_i} \comma \quad \bar{i}_{-}\sim (\barz -\barz_i )^{1-\eta_i}/ (1-2\eta_i) \period \label{CL-AntiHolAsympt}
}
Here, we normalized $i_{\pm}$ so that they satisfy the normalization conditions, $\langle i_{+},i_{-}\rangle=\langle \bar{i}_{+},\bar{i}_{-}\rangle=1$.
In terms of $i_{\pm}$ and $\bar{i}_{\pm}$ defined above, the solutions to \eqref{CL-AuxLinProb} which appears in the formula \eqref{CL-Reconst} can be expressed as follows.
\beq{
&\psi_k = \langle \psi_k \comma  i_- \rangle i_{+}-\langle \psi_k \comma  i_+ \rangle i_- \comma \\
&\bar{\psi}_k = \langle \bar{\psi}_k \comma  \bar{i}_- \rangle \bar{i}_{+}-\langle \bar{\psi}_k \comma  \bar{i}_+ \rangle \bar{i}_- \comma 
}
These expressions are useful to determine the asymptotic behavior of $\psi_k$ and $\bar{\psi}_k$. Since $\eta_i $ is smaller\fn{Recall that $\eta_i$ is smaller than $1/2$ owing to the condition \eqref{phys-reg}.} than $1-\eta_i$, $\psi_k$ and $\bar{\psi}_k$ can be approximated near $z_i$ as
\beq{
\psi_k {\sim} \langle \psi_k \comma  i_- \rangle (z-z_i)^{\eta _i}\comma\quad
\bar{\psi}_k {\sim} \langle \bar{\psi}_k \comma  \bar{i}_- \rangle (\barz -\barz _i)^{\eta _i}\period
}
Thus, the asymptotic behavior of $\phi$ can be determined as
\beq{
\phi \sim -2\eta_i \log |z-z_i|-\log \left(  \langle \psi_1 \comma  i_-\rangle \langle \bar{\psi}_1 \comma  \bar{i}_-\rangle -\langle \psi_2 \comma  i_-\rangle \langle \bar{\psi}_2 \comma  \bar{i}_-\rangle \right)-\frac{1}{2}\log\lambda \period
}

We have not hitherto considered seriously the single-valuedness of $e^{-\phi}$.
Thus, let us next discuss what kind of constraints the single-valuedness imposes.
Although we will henceforth consider only the three-point functions, the discussion in this subsection can be generalized to the higher-point functions.
First, consider the single-valuedness around $z_1$. To be single-valued, or equivalently monodromy-free, around $z_1$, $e^{-\phi}$ must be of the following form:
\beq{
e^{-\phi}/\sqrt{\lambda} = A  1_{+} \bar{1}_{+} - A^{-1} 1_{-} \bar{1}_{-}\period \label{CL-SingleValue}
}  
In other words, $\psi_1$ and $\bar{\psi}_1$ must be proportional to $1_{+}$ and $\bar{1}_+$, and $\psi_2$ and $\bar{\psi}_2$ must be proportional to $1_{-}$ and $\bar{1}_{-}$. 
For the reality of $e^{-\phi}$, we should take $A$ to be real.
Now, let us re-express \eqref{CL-SingleValue} in terms of the solution around $z_2$\fn{The monodromy-free condition around $z_3$ is trivial since the contour around $z_3$ can be decomposed into a sum of contours around $z_1$ and $z_2$.}. Using the following relation,
\beq{
1_{+} = \langle 1_{+} \comma 2_{-}\rangle 2_{+}-\langle 1_{+} \comma 2_{+}\rangle 2_{-} \text{  {\it etc.}}
}
we obtain
\beq{
e^{-\phi}/\sqrt{\lambda}=&  \left( A \langle 1_{+} \comma 2_{-}\rangle \langle \bar{1}_{+} \comma \bar{2}_{-}\rangle - A^{-1}\langle 1_{-} \comma 2_{-}\rangle \langle \bar{1}_{-} \comma \bar{2}_{-}\rangle\right) 2_{+} \bar{2}_{+} \nn\\
&+ \left( A \langle 1_{+} \comma 2_{+}\rangle \langle \bar{1}_{+} \comma \bar{2}_{+}\rangle - A^{-1}\langle 1_{-} \comma 2_{+}\rangle \langle \bar{1}_{-} \comma \bar{2}_{+}\rangle\right) 2_{-} \bar{2}_{-}\nn\\
&-\left( A \langle 1_{+} \comma 2_{-}\rangle \langle \bar{1}_{+} \comma \bar{2}_{+}\rangle - A^{-1}\langle 1_{-} \comma 2_{-}\rangle \langle \bar{1}_{-} \comma \bar{2}_{+}\rangle\right) 2_{+} \bar{2}_{-}\nn\\
&-\left( A \langle 1_{+} \comma 2_{+}\rangle \langle \bar{1}_{+} \comma \bar{2}_{-}\rangle - A^{-1}\langle 1_{-} \comma 2_{+}\rangle \langle \bar{1}_{-} \comma \bar{2}_{-}\rangle\right) 2_{-} \bar{2}_{+}\period
}
$e^{-\phi}$ is single-valued if and only if the last two terms in the above equation vanishes. Therefore, we obtain the following constraints,
\beq{
&A \langle 1_{+} \comma 2_{-}\rangle \langle \bar{1}_{+} \comma \bar{2}_{+}\rangle - A^{-1}\langle 1_{-} \comma 2_{-}\rangle \langle \bar{1}_{-} \comma \bar{2}_{+}\rangle =0\comma\label{CL-SingleValueConst1}\\
&A \langle 1_{+} \comma 2_{+}\rangle \langle \bar{1}_{+} \comma \bar{2}_{-}\rangle - A^{-1}\langle 1_{-} \comma 2_{+}\rangle \langle \bar{1}_{-} \comma \bar{2}_{-}\rangle =0 \period\label{CL-SingleValueConst2}
}
Solving the above two equations \eqref{CL-SingleValueConst1} and \eqref{CL-SingleValueConst2}, we arrive at the following expression:
\beq{
\frac{e^{-\phi}}{\sqrt{\lambda}} &= 
\sqrt{\left| \frac{\langle 1_{-} \comma 2_{-}\rangle \langle 1_{-} \comma 2_{+}\rangle}{\langle 1_{+} \comma 2_{+}\rangle \langle 1_{+} \comma 2_{-}\rangle}\right|}
1_{+}\bar{1}_{+}
-
\sqrt{\left| \frac{\langle 1_{+} \comma 2_{+}\rangle \langle 1_{+} \comma 2_{-}\rangle}{\langle 1_{-} \comma 2_{-}\rangle \langle 1_{-} \comma 2_{+}\rangle}\right|}
1_{-}\bar{1}_{-}\period\label{asym-1p1m}
}
From the single-valuedness condition around $z_3$, we can derive an equation similar to \eqref{asym-1p1m}, but with $2_{\pm}$ replaced with $3_{\pm}$. Combining these two expressions, the following symmetric expression can be written down:
\beq{
\frac{e^{-\phi}}{\sqrt{\lambda}}  =&
\left| \frac{\langle 1_{-} \comma 2_{-}\rangle \langle 1_{-} \comma 2_{+}\rangle}{\langle 1_{+} \comma 2_{+}\rangle \langle 1_{+} \comma 2_{-}\rangle}\frac{\langle 3_{-} \comma 1_{-}\rangle \langle 3_{+} \comma 1_{-}\rangle}{\langle 3_{+} \comma 1_{+}\rangle \langle 3_{-} \comma 1_{+}\rangle}\right|^{\frac{1}{4}}
1_{+}\bar{1}_{+}\nn\\
&
-\left| \frac{\langle 1_{+} \comma 2_{+}\rangle \langle 1_{+} \comma 2_{-}\rangle}{\langle 1_{-} \comma 2_{-}\rangle \langle 1_{-} \comma 2_{+}\rangle}\frac{\langle 3_{+} \comma 1_{+}\rangle \langle 3_{-} \comma 1_{+}\rangle}{\langle 3_{-} \comma 1_{-}\rangle \langle 3_{+} \comma 1_{-}\rangle}\right|^{\frac{1}{4}}
1_{-}\bar{1}_{-}\period\label{CL-WronskianAround1}
}
The expression in terms of $2_{\pm}$ or in terms of $3_{\pm}$ can be obtained by the permutation of $1$, $2$ and $3$ as follows:
\beq{
\frac{e^{-\phi}}{\sqrt{\lambda}}  =&
\left| \frac{\langle 2_{-} \comma 3_{-}\rangle \langle 2_{-} \comma 3_{+}\rangle}{\langle 2_{+} \comma 3_{+}\rangle \langle 2_{+} \comma 3_{-}\rangle}\frac{\langle 1_{-} \comma 2_{-}\rangle \langle 1_{+} \comma 2_{-}\rangle}{\langle 1_{+} \comma 2_{+}\rangle \langle 1_{-} \comma 2_{+}\rangle}\right|^{\frac{1}{4}}
2_{+}\bar{2}_{+}\nn\\
&-
\left| \frac{\langle 2_{+} \comma 3_{+}\rangle \langle 2_{+} \comma 3_{-}\rangle}{\langle 2_{-} \comma 3_{-}\rangle \langle 2_{-} \comma 3_{+}\rangle}\frac{\langle 1_{+} \comma 2_{+}\rangle \langle 1_{-} \comma 2_{+}\rangle}{\langle 1_{-} \comma 2_{-}\rangle \langle 1_{+} \comma 2_{-}\rangle}\right|^{\frac{1}{4}}
2_{-}\bar{2}_{-}\comma\label{CL-WronskianAround2}\\
\frac{e^{-\phi}}{\sqrt{\lambda}}  =&
\left| \frac{\langle 3_{-} \comma 1_{-}\rangle \langle 3_{-} \comma 1_{+}\rangle}{\langle 3_{+} \comma 1_{+}\rangle \langle 3_{+} \comma 1_{-}\rangle}\frac{\langle 2_{-} \comma 3_{-}\rangle \langle 2_{+} \comma 3_{-}\rangle}{\langle 2_{+} \comma 3_{+}\rangle \langle 2_{-} \comma 3_{+}\rangle}\right|^{\frac{1}{4}}
3_{+}\bar{3}_{+}\nn\\
&-
\left| \frac{\langle 3_{+} \comma 1_{+}\rangle \langle 3_{+} \comma 1_{-}\rangle}{\langle 3_{-} \comma 1_{-}\rangle \langle 3_{-} \comma 1_{+}\rangle}\frac{\langle 2_{+} \comma 3_{+}\rangle \langle 2_{-} \comma 3_{+}\rangle}{\langle 2_{-} \comma 3_{-}\rangle \langle 2_{+} \comma 3_{-}\rangle}\right|^{\frac{1}{4}}
3_{-}\bar{3}_{-}\period\label{CL-WronskianAround3}
}
From \eqref{CL-WronskianAround1}-\eqref{CL-WronskianAround3}, the parameters $C_i$ can be easily read off as
\beq{
&C_i =-\frac{1}{2}\log \lambda + \frac{1}{4}\left( \sum_{\epsilon = \pm,  j\neq i } \log \left| \langle i_{+}\comma j_{\epsilon}\rangle\right| -\sum_{\epsilon = \pm, j\neq i} \log \left| \langle i_{-}\comma j_{\epsilon}\rangle\right| \right)\period\label{CL-C1}
}
Thus, the semiclassical three-point functions can be computed once we know the values of the Wronskians $\langle i_{\pm}\comma j_{\pm}\rangle$. 

\section{Determination of Wronskians}
\label{Sec-Wr}
%%%%%%
%In this section we add "WR-" to the head of labels
%%%%%%

Having seen that the Wronskians are fundamental objects for the evaluation of the semiclassical three-point functions, we shall evaluate such Wronskians in this section.
First we show that certain products of Wronskians can be determined from the triviality of the monodromy at infinity.
To obtain the individual Wronskians, we next consider a one-parameter deformation of the auxiliary linear problem and the Wronskians.
Then, we determine the analyticity of the Wronskians and set up the Riemann-Hilbert problem.
Finally, solving the Riemann-Hilbert problem, we compute the parameters $C_i$ in \eqref{CL-C1} and see that $C_i$'s thus computed coincide with the known result derived in \cite{ZZ,HJ,HMW}.

\subsection{Monodromy relation}
\label{Sec-Wr-Mon}

Let us consider the monodromy of solutions to the first equation of \eqref{CL-AuxLinProb}.
Since we are considering the solutions on the sphere, only noncontractable cycles are cycles around punctures (vertex operators). The eigenvalues of each monodromy matrix are determined purely by the local behavior of the solutions and therefore can be expressed purely by the parameter of the corresponding vertex operator.
\beq{
&\Omega_i \sim \pmatrix{cc}{e^{ip_i}&0\\0&e^{-ip_i}}\comma \label{WR-DiagonalMonodromy}
}
where $p_i$ is the Liouville momentum defined by $p_i \equiv 2\pi \eta_i\period$

Although $\Omega_i$'s can be separately diagonalized as \eqref{WR-DiagonalMonodromy}, they cannot be diagonalized simultaneously, owing to the nontrivial global behavior of the solution to the auxiliary linear problem.
Nevertheless, the simple relation, $\Omega_1\Omega_2\Omega_3=1$, which expresses the triviality of the monodromy at infinity, is enough to restrict the form of $\Omega_i$ sufficiently, although not completely.
To see this, let us take a basis of solutions in which $\Omega_1$ is diagonal,
\beq{
\Omega_1 = \pmatrix{cc}{e^{ip_1}&0\\0&e^{-ip_1}}\comma
}
and express the matrix $\Omega_2$ in this basis as
\beq{
\Omega_2 =\pmatrix{cc}{a&b\\c&d}\period
}
Since $\Omega_2$ has eigenvalues $\exp(\pm ip_2)$, $a$-$d$ must satisfy the following relations.
\beq{
a+d = 2\cos p_2 \comma\quad ad-bc=1\period \label{WR-MonEigen2}
}
Next, let us consider $\Omega_3$. Owing to the relation, $\Omega_1\Omega_2\Omega_3=1$, $\Omega_3$ can be expressed in terms of $a$-$d$ as follows.
\beq{
\Omega_3 =\Omega_2^{-1}\Omega_{1}^{-1}=\pmatrix{cc}{e^{-ip_1}d&-e^{ip_1}b\\-e^{-ip_1}c&e^{ip_1}a}\period
}
Then, using the fact $\Omega_3$ has eigenvalues $\exp (\pm ip_3)$, we can obtain one more constraint.
\beq{
e^{-ip_1}d+e^{ip_1}a=2\cos p_3 \label{WR-MonEigen3}
}
It is easy to see that \eqref{WR-MonEigen2} and \eqref{WR-MonEigen3} completely determines $a$ and $d$, while only the product $bc$ can be determined\fn{For a detail of the derivation, see \cite{KK1}.}. With these explicit form of the monodromy matrix, we can compute various Wronskians between $i_{\pm}$, which are the eigenvectors of the monodromy matrices. However, because of the aforementioned ambiguity of $b$ and $c$, only the following products of Wronskians can be determined:
\beq{
&\sl{1_+\comma 2_+}\sl{1_-\comma 2_-} = \frac{\sin \frac{p_1+p_2-p_3}{2}\sin \frac{p_1+p_2+p_3}{2}}{\sin p_1 \sin p_2}\comma \label{WR-Product1}\\
&\sl{2_+\comma 3_+}\sl{2_-\comma 3_-} = \frac{\sin \frac{-p_1+p_2+p_3}{2}\sin \frac{p_1+p_2+p_3}{2}}{\sin p_2 \sin p_3}\comma\label{WR-Product2}\\
&\sl{3_+\comma 1_+}\sl{3_-\comma 1_-} = \frac{\sin \frac{p_1-p_2+p_3}{2}\sin \frac{p_1+p_2+p_3}{2}}{\sin p_3 \sin p_1}\comma\label{WR-Product3}\\
&\sl{1_+\comma 2_-}\sl{1_-\comma 2_+} = \frac{\sin \frac{p_1-p_2-p_3}{2}\sin \frac{p_1-p_2+p_3}{2}}{\sin p_1 \sin p_2}\comma \label{WR-Product4}\\
&\sl{2_+\comma 3_-}\sl{2_-\comma 3_+} = \frac{\sin \frac{-p_1+p_2-p_3}{2}\sin \frac{p_1+p_2-p_3}{2}}{\sin p_2 \sin p_3}\comma\label{WR-Product5}\\
&\sl{3_+\comma 1_-}\sl{3_-\comma 1_+} = \frac{\sin \frac{-p_1-p_2+p_3}{2}\sin \frac{-p_1+p_2+p_3}{2}}{\sin p_3 \sin p_1}\period\label{WR-Product6}
}

To compute the semiclassical three-point functions, we need to compute the individual Wronskians, not just their products. For this purpose, let us introduce the following one-parameter deformation of the auxiliary linear problem:
\beq{
\left( \del ^2 +\frac{1}{\xi ^2} T(z)\right) \psi =0\period \label{WR-SpectralDefAux}
}
In what follows, we will call $\xi$ the \textit{spectral parameter} following the convention in integrable systems. \eqref{WR-SpectralDefAux} coincide in form with the Schr\"{o}dinger equation\fn{Later, utilizing this analogy, we will discuss the WKB expansion of \eqref{WR-SpectralDefAux}.} if we identify $\xi$ with $\hbar$.
The asymptotic behavior of the solutions $i_\pm$ \eqref{CL-HolAsympt} and \eqref{CL-AntiHolAsympt} is modified as
\beq{
&i_{+} \sim (z-z_i)^{\eta_i(\xi)} \comma \quad i_{-}\sim (z-z_i)^{1-\eta_i(\xi)}/ (1-2\eta_i(\xi)) \comma \label{WR-HolAsymptMod}\\
&\bar{i}_{+}\sim (\barz -\barz_i)^{\eta_i(\xi)} \comma \quad \bar{i}_{-}\sim (\barz -\barz_i )^{1-\eta_i(\xi)}/ (1-2\eta_i(\xi)) \comma \label{WR-AntiHolAsymptMod}
}
where
\beq{
\eta_i(\xi)\equiv \frac{1}{2}-\sqrt{\frac{1}{4}-\frac{\eta_i (1-\eta_i)}{\xi^2}}\period
}
Thus, all the formulae derived in this subsection, in particular the formulae for the products of Wronskians \eqref{WR-Product1}-\eqref{WR-Product6}, are valid under the following replacement:
\beq{
p_i \rightarrow p_i(\xi)\equiv 2 \pi \eta_i(\xi)\period
}
Below, we often omit writing the dependence on $\xi$ and denote $p_i(\xi)$ simply by $p_i$.

\subsection{Poles of Wronskians}
Having introduced the one-parameter deformation, we will next discuss the analytic properties, namely poles and zeros, of the Wronskians with respect to $\xi$. As \eqref{WR-Product1}--\eqref{WR-Product6} shows, the products of the Wronskians have poles at $\sin p_i=0$ and zeros at $\sin \left(\sum_i \epsilon_i p_i/2 \right)=0$, where $\epsilon_i$ takes $+1$ or $-1$. Thus, our practical task is to allocate these poles and zeros between two Wronskians.

Let us first discuss poles of the Wronskians. For simplicity, we will only consider poles at $\sin p_1 =0$. Generalization to other poles is straightforward.
At $\sin p_1=0$, two eigenvalues of $\Omega_1$, $e^{+ip_1}$ and $e^{-ip_1}$ become both $+1$ or both $-1$. This, however, does not mean that $\Omega_1$ is proportional to the unit matrix at such points: If $\Omega_1$ is proportional to the unit matrix, $\Omega_2$ must satisfy $\Omega_2 = \pm \Omega_{3}^{-1}$ owing to the monodromy relation $\Omega_1 \Omega_2 \Omega_3=\mathbf{1}$. However, for generic $p_2$ and $p_3$, there is no reason for this to be satisfied since $p_1$, $p_2$ and $p_3$ can be chosen completely independently. Therefore, the only remaining possibility is that $\Omega_1$ becomes a Jordan-block:
\beq{
\Omega_1 \sim \pm \pmatrix{cc}{1&c\\0&1}\quad \text{at }\sin p_1 =0\period
}
This means two eigenvectors of $\Omega_1$ degenerate at $\sin p_1=0$.

To see what happens at $\sin p_1 =0$ more explicitly, let us study the behavior of $1_{\pm}$ near $z_1$. Owing to the normalization conditions \eqref{WR-HolAsymptMod}, the expansion of $1_{\pm}$ around $z_1$ is in general given by
\beq{
&1_{+} = (z-z_1)^{\eta_1(\xi)}\left(1+ a_{+}^{(1)} (z-z_1) +a_{+}^{(2)} (z-z_1)^2 +\cdots \right)\comma\label{1+exp}\\
&1_{-}= \frac{(z-z_1)^{1-\eta_1(\xi)}}{1-2\eta_1(\xi)}\left( 1+ a_{-}^{(1)} (z-z_1) +a_{-}^{(2)} (z-z_1)^2 +\cdots \right)\comma\label{1-exp}
}
where $a_{\pm}^{(n)}$ are functions of $\xi$. An important observation here is that, although $1_{+}$ and $1_{-}$ degenerate at $\sin p_1 =0$, their leading terms, $(z-z_1)^{\eta_1}$ and $(z-z_1)^{1-\eta_1}/(1-2\eta_1)$, do not degenerate in general. This apparent contradiction can be resolved if and only if one of $1_{\pm}$ appears in the expansion of the other with a divergent coefficient. More precisely, we expect one of the following two situations is realized at $\sin p_1=0$:
\beq{
&1_{+}=(z-z_1)^{\eta_1(\xi)}+\cdots + A(\xi ) 1_{-}+\cdots\comma\label{1+appear}\\
&A(\xi)\to \infty \quad \text{at }\sin p_1 =0\comma\nn
}
or
\beq{
&1_{-}=\frac{(z-z_1)^{1-\eta_1(\xi)}}{1-2\eta_1(\xi)}+\cdots + B(\xi ) 1_{+}+\cdots\comma\label{1-appear}\\
&B(\xi)\to \infty \quad \text{at }\sin p_1 =0\period\nn
}
The first case \eqref{1+appear} is realized if $1_{+}$ is bigger than $1_-$ in the neighborhood of $z_1$, namely if $\eta_1(\xi)$ is smaller than $1-\eta_1(\xi)$. On the other hand, the second case \eqref{1-appear} is realized if $1_{+}$ is smaller than $1_-$ around $z_1$ and $\eta_1(\xi)$ is bigger than $1-\eta_1(\xi)$. From this simple analysis, we can conclude that, among two solutions $i_{\pm}$, the solution which is bigger around $z_i$ diverges at $\sin p_i=0$ whereas the solution which is smaller is finite. This leads to the following rules for determination of poles of the Wronskians:
\beq{
\text{At }\sin p_i=0 \qquad \left[\begin{aligned}
&\text{if }\eta_i(\xi) <1/2 \Rightarrow \sl{i_{+}\comma \bullet}= \infty \comma \quad\sl{i_{-}\comma  \bullet}\neq \infty \comma\\
&\text{if }\eta_i(\xi) \geq 1/2 \Rightarrow \sl{i_{+}\comma \bullet}\neq \infty \comma \quad \sl{i_{-}\comma \bullet}= \infty \period
\end{aligned}\right.
}
Note that, at the boundary value $\eta_i(\xi)=1/2$, the Wronskians containing $i_{-}$ diverge owing to the explicit factor $1/(1-2\eta_i(\xi))$ contained in \eqref{WR-HolAsymptMod}.
\subsection{Zeros of Wronskians}
Having determined the pole structures, let us next discuss zeros of the Wronskians. The determination of zeros is substantially more difficult than the determination of poles since zeros are determined by the global properties on the Riemann surface while the poles are determined solely by the local properties around the punctures. As shown in the previous works \cite{GMN2,AMSV,JW,KK1,CT}, the WKB curve is one of the central tools to explore such global properties. However, as its name suggests, the WKB curve is useful only when we can approximate the solutions to the auxiliary linear problem by the leading term in the WKB expansion. For this reason, it is not powerful enough to fully determine the zeros of the Wronskians in the whole spectral parameter region. In this subsection, we shall introduce an appropriate generalization of the WKB curve, to be called the {\it exact WKB curve}, which allows us to determine zeros of the Wronskian on the whole parameter plane.
\subsubsection{WKB approximation and WKB curves}
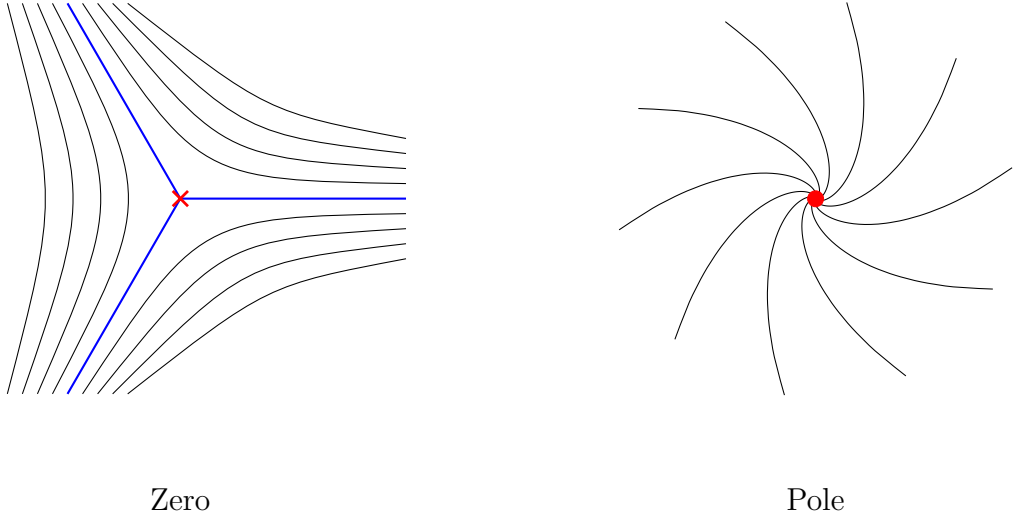
\begin{figure}
\begin{minipage}{0.5\hsize}
\centering
\begin{tikzpicture}
\draw[blue,thick] (0,0) -- (3,0)
(0,0) -- (-3/2,{3/2*sqrt(3)})
(0,0) -- (-3/2,{-3/2*sqrt(3)});
\foreach \x in {2,4,...,8}
{\draw ({-3/2+0.1*\x},{3/2*sqrt(3)}) .. controls ({0.15*\x -0.05},{0.15*\x -0.05}) .. (3,{0.1*\x})
({-3/2+0.1*\x},{-3/2*sqrt(3)}) .. controls ({0.15*\x -0.05},{-0.15*\x +0.05}) .. (3,{-0.1*\x})
({-3/2-0.1*\x},{3/2*sqrt(3)}) .. controls ({-(0.15*\x -0.05)*sqrt(2)},0) .. ({-3/2-0.1*\x},{-3/2*sqrt(3)});
}
\node at (0,0) {\Cross};
\node at (0,-4) {Zero};
\end{tikzpicture}
\end{minipage}
\begin{minipage}{0.5\hsize}
\centering
\begin{tikzpicture}
\foreach \x in {1,...,10}{
\draw[scale=0.55,domain={-pi}:{pi/2},smooth,variable=\t]
plot ({exp(\t)*cos((0.5*\t + \x*pi/5) r)},{exp(\t)*sin((0.5*\t + \x*pi/5) r)});
}
\filldraw[red] (0,0) circle (3pt);
\node at (0,-4) {Pole};
\end{tikzpicture}
\end{minipage}
\caption{Structures of the WKB curves around zeros and poles.
There are three WKB curves radiating from a zero, while infinitely many WKB curves emanate from a pole.
The exact WKB curves, to be discussed later, also have similar structures.}
\label{fig-WKB}
\end{figure}

Before introducing the generalized version, let us briefly review the ordinary WKB curves defined in \cite{GMN2}. When $\xi$ is sufficiently close to zero, we can construct approximate solutions to the auxiliary linear problem by the WKB expansion. For instance, the approximate solutions around the singularity $z_i$ can be constructed as follows:
\beq{
\psi \sim \exp \left(\pm \frac{1}{\xi}\int_{z_i+\epsilon_i}^{z} \sqrt{-T}dz \right)\comma\label{leadwkb}
}
where $\epsilon_i(\ll 1)$ is the regularization parameter.
\eqref{leadwkb} gives two different approximate solutions depending on the choice of the sign. One of the solutions is the {\it small solution}, which decreases exponentially as it approaches $z_i$. The other is the {\it big solution}, which increases exponentially as it approaches $z_i$. Now let us define the WKB curves as the curves along which the phase of the leading term \eqref{leadwkb} in the WKB expansion is constant. More explicitly, they are given by
\beq{
\Im \left( \frac{\sqrt{-T}}{\xi}dz\right) =0\period\label{def-wkb}
}
By analyzing \eqref{def-wkb}, we can confirm that there are three different structures of the WKB curves. First, at generic points on the worldsheet, the WKB curves are completely smooth and non-intersecting. Second, at a puncture, the WKB curves radiate in all direction from the puncture. Third, at a zero of $T(z)$, there are three special WKB curves\fn{These special WKB curves are called {\it Stokes lines} in the literature.} which radiate from the zero and separate three different configurations of the WKB curves. For a detail, see \figref{fig-WKB}.
Along the WKB curve, the magnitude of the leading term in the WKB expansion \eqref{leadwkb} monotonically increases or monotonically decreases until they reach a zero or a pole of $T(z)$. Thus, if two punctures $z_i$ and $z_j$ are connected by a WKB curve and the spectral parameter $\xi$ is sufficiently small, the small solution defined around $z_i$, to be denoted $s_i$, will grow exponentially as it approaches the other puncture $z_j$. This means that the small solution $s_i$ behaves like the big solution around $z_j$. Therefore it will be linearly independent of $s_j$, the small solution defined at $z_j$. Consequently, the Wronskian between these two small solutions $\sl{s_i\comma s_j}$ will be nonzero. Applying this logic, we conclude that the Wronskians $\sl{i_{\pm}\comma j_{\pm}}$ are nonzero if the following three conditions are satisfied: First, two punctures $z_i$ and $z_j$ are connected by the WKB curves. Second, two eigenvectors $i_{\pm}$ and $j_{\pm}$ are both small solutions. Third, the approximation by the leading WKB solutions \eqref{leadwkb} is sufficiently good. 

\subsubsection{Exact WKB curves}
Evidently, the above analysis is valid only in a restricted region on the spectral parameter plane where the approximation by the leading term in the WKB expansion is reliable. To understand the structure of zeros on the whole spectral parameter plane, we need to generalize the notion of WKB curves.

To motivate our definition of the exact WKB curves, let us first make a small detour and discuss the general structure of the WKB expansions.
If we express the solution to the auxiliary linear problem \eqref{WR-SpectralDefAux} as
\beq{
&\psi(z,\xi)=\exp \left( \int^z_{z_0} P(z',\xi) dz' \right) \comma \quad P(z,\xi)=\sum^{\infty}_{n=-1} \xi^n  P_{n}(z) \comma
}
we obtain the following Riccati equation for $P(z;\xi)$:
\beq{
P^2+\partial P + \frac{1}{\xi^2} T(z)=0 \period
}
Since the left hand side of this equation must vanish at each order of $\xi$ expansion, $P_n$ can be determined recursively,
\beq{
P_{-1}^2=-T\comma\quad P_n=\frac{-1}{2 P_{-1}} \left(\sum^{n-1}_{m=0}P_m P_{n-m-1} +\del P_{n-1}\right)\comma \quad n\geq 0\period
}
Then, dividing $P$ into the odd powers and the even powers of $\xi$,
\beq{
P=P_{\rm odd}+P_{\rm even}\comma \quad P_{\rm odd}=\sum^\infty_{n=0} \xi^{2n-1} P_{2n-1}\comma \quad P_{\rm even}=\sum^\infty_{n=0} \xi^{2n} P_{2n}\comma
}
we obtain the following relation,
\beq{
P_{\rm even}=-\frac{1}{2} \del \log P_{\rm odd}\period\label{evenodd}
}
Using \eqref{evenodd}, we can express the solutions only in terms of $P_{\rm odd}$.
In summary, the WKB expansion of two independent solutions to the auxiliary linear problem can be expressed in the following simple form:
\beq{
\psi^{(\pm)}_{\rm WKB}=\frac{1}{\sqrt{P_{\rm WKB}(z,\xi)}}\exp \left( \pm \int^z_{z_0} P_{\rm WKB}(z',\xi) dz' \right)\comma\label{WKBP}
}
where $P_{\rm WKB}$ in \eqref{WKBP} denotes $P_{\rm odd}$ in the preceding equations. 

Motivated by the simple form of \eqref{WKBP}, let us express the exact solutions to the auxiliary linear problem as
\beq{
\psi = \frac{1}{\sqrt{2P_{\rm ex}(z;\xi)}}\exp\left( - \int_{z_0}^{z} P_{\rm ex}(z;\xi) dz \right)\period \label{exactP}
}
Although \eqref{WKBP} and \eqref{exactP} are almost identical in form, there is a notable difference: While $P_{\rm WKB}$ is given only by an asymptotic series with respect to $\xi$ and is ambiguous in a non-perturbative sense, $P_{\rm ex}$ is defined unambiguously as it is directly defined by the exact solution, $\psi$. Of course, if we expand $P_{\rm ex}$ perturbatively with respect $\xi$, we will get the same expansion as $P_{\rm WKB}$. Therefore, $P_{\rm ex}$ can be regarded as a non-perturbative completion of $P_{\rm WKB}$.
One of the virtues of the expression \eqref{exactP} is that we can easily construct another solution satisfying $\sl{\psi \comma \tilde{\psi}}=1$ as
\beq{
\tilde{\psi} = \frac{1}{\sqrt{2P_{\rm ex}(z;\xi)}}\exp\left( + \int_{z_0}^{z} P_{\rm ex}(z;\xi) dz \right)\period \label{exactP2}
} 
Using \eqref{exactP}, let us now discuss the generalization of the WKB curves.
The quantity we used to define the original WKB curves, $\sqrt{-T}/\xi$, is the leading term in the expansion of $P_{\rm WKB}$. Therefore, the most natural generalization of the WKB curves would be to use $P_{\rm ex}$, which is a non-perturbative completion of $P_{\rm WKB}$ and define the curves by
\beq{
\Im \left( P_{\rm ex}(z;\xi) dz\right)=0\period
}
However, this definition is ambiguous since a different choice of the exact solution $\psi$ leads to a different $P_{\rm ex}$ and thus to different curves. To fix such ambiguities, we define the generalization of the WKB curves, to be called {\it exact WKB curves}, separately for each puncture:
\begin{itemize}
\item[] {\it Exact WKB curves.} Exact WKB curves for the puncture $z_i$, to be denoted eWKB$_{(i)}$, are defined by
 \beq{
 \Im \left( P^{(i)}_{\rm ex}(z;\xi) dz\right) =0\comma\label{exWKB}
 }
where $P^{(i)}_{\rm ex}$ is the exponential factor for the smaller eigenvector among $i_{\pm}$. More explicitly, it is defined by
\beq{
s_i = \frac{1}{\sqrt{2P_{\rm ex}^{(i)}(z;\xi)}}\exp\left( - \int_{z_i+\epsilon_i}^{z} P_{\rm ex}^{(i)}(z;\xi) dz \right)\comma\label{si}
}
where $s_i$ denotes the smaller solution among $i_{\pm}$ and $\epsilon_i(\ll 1)$ is the parameter we introduced for the regularization.
\end{itemize}
Let us now make several comments. First, it is easy to see that this definition of the exact WKB curves reduces to that of the ordinary WKB curves when $\xi$ is close to zero. Second, if we flip the sign in the exponent
of \eqref{si} as
\beq{
b_i \equiv  \frac{1}{\sqrt{2P_{\rm ex}^{(i)}(z;\xi)}}\exp\left( + \int_{z_i+\epsilon_i}^{z} P_{\rm ex}^{(i)}(z;\xi) dz \right)\comma\label{bi}
}
we obtain another solution $b_i$, which is big near the puncture $z_i$ and satisfies $\sl{s_i \comma b_i}=1$. The solution $b_i$, constructed in this way, is not guaranteed to be an eigenvector of $\Omega_i$. The other eigenvector is in general given by some linear combination of $s_i$ and $b_i$, $b_i + c s_i$. However, as $s_i$ is exponentially smaller than $b_i$ near $z_i$, $b_i$ coincide, up to exponentially small corrections, with the eigenvector in the neighborhood of $z_i$.

When $\xi$ is not close to zero, $P_{\rm ex}^{(i)}$ and $P_{\rm ex}^{(j)}$ will in general differ by the terms non-perturbative with respect to $\xi$. Therefore eWKB$_{(i)}$ and eWKB$_{(j)}$ will be different. Nevertheless, in a sufficient neighborhood of $z_i$ or $z_j$, we expect eWKB$_{(i)}$ and eWKB$_{(j)}$ to be identical. To see this, let us consider the behavior of the small solution $s_i$ along eWKB$_{(i)}$. Along eWKB$_{(i)}$, the phase of the exponential factor in $s_i$ is constant and its magnitude monotonically increases\fn{Strictly speaking, the small eigenvector \eqref{si} also contains a prefactor in front of the exponential. This prefactor, however, does not play a significant role in our discussion since it drops out if we consider the ratio of two solutions $s_i/b_i$. It is in fact sufficient to know the ratio in order to identify the small solution and the big solution.} until it reaches some endpoint. Then, if the punctures $z_i$ and $z_j$ are connected by 
 some curves in eWKB$_{(i)}$, $s_i$ will grow exponentially as it approaches $z_j$ along the curves, which means that $s_i$ behaves like the big solution around $z_j$. Since requiring a solution to be exponentially big around a puncture does not uniquely specify the solution, $s_i$ is in general expressed as $s_i\propto b_j +c s_j$, where $c$ is some unknown constant. However, since $s_j$ is exponentially smaller than $b_j$ around $z_j$, we can neglect the second term $c s_j$ and approximate $s_i$ in the neighborhood of $z_j$ as
\beq{
s_i \propto b_j = \frac{1}{\sqrt{2P_{\rm ex}^{(j)}(z;\xi)}}\exp\left( + \int_{z_j+\epsilon_j}^{z} P_{\rm ex}^{(j)}(z;\xi) dz \right)\comma\label{nearzj}
} 
up to exponentially small corrections. From \eqref{nearzj}, we can show that $P_{\rm ex}^{(i)}\sim -P_{\rm ex}^{(j)}$ near the puncture $z_j$. Then, since the exact WKB curves do not depend on the overall sign of $P_{\rm ex}^{(i)}$, we conclude eWKB$_{(i)}$ and eWKB$_{(j)}$ coincide in the vicinity of $z_j$.

Let us now use the exact WKB curves to determine the analyticities of the Wronskians. Following exactly the same logic as in the case of the ordinary WKB curves, we can conclude that the Wronskian involving two small solutions $s_i$ and $s_j$ must be nonzero if two punctures $z_i$ and $z_j$ are connected by some exact WKB curves. However, unlike the ordinary WKB curves, it is difficult to determine the configuration of the exact WKB curves since the explicit form of $P^{(i)}_{\rm ex}$ cannot be obtained unless we solve the auxiliary linear problem explicitly. Nevertheless, we shall show below that the local properties of the exact WKB curves, in particular the ``number densities" of the curves emanating from punctures, turn out to give enough restrictions to determine the full topology (connectivity) of the curves.
To see this, let us first give a definition of the number density of the exact WKB curves. Since the exact WKB curves are defined as curves on which $\Im \int P_{\rm ex}^{(i)}dz$ is constant, the number density of the exact WKB curves emanating from $z_i$ can be counted by measuring how $\Im \int P_{\rm ex}^{(i)} dz$ changes as we go around the puncture. This leads to the following formula:
\beq{
N_i\equiv \frac{1}{2\pi}\oint_{\mathcal{C}_i}|\Im \, P_{\rm ex}^{(i)} dz|\comma\label{numex}
}
where the integration contour $\mathcal{C}_i$ is an infinitesimal circle encircling the puncture $z_i$ counterclockwise.
From the asymptotic behavior of $i_{\pm}$ \eqref{WR-HolAsymptMod}, we can determine the behavior of $P_{\rm ex}^{(i)}$ as follows:
\beq{
P_{\rm ex}^{(i)} \sim \pm \left(\eta_i(\xi)-\frac{1}{2}\right) \frac{1}{z-z_i} \quad \text{as }z\to z_i \label{appex}
}
where $+$ or $-$ sign is chosen depending on which of $i_{\pm}$ is small. Then
$N_i$ can be computed explicitly as
\beq{
N_i \equiv \left|\Re \left(  \eta_i (\xi )-\frac{1}{2}\right)\right|\period\label{defni}
}
Next, we shall derive two important properties of the eWKB$_{(i)}$'s which will be necessary to determine their configurations. Let us first show the following property:
\begin{itemize}
\item[]{\it Non-contractibility}. All the exact WKB curves which start and end at the same puncture are non-contractible. In other words, such curves go around a different puncture at least once.
\end{itemize}
To understand this, recall that the Wronskians between small solutions should be nonzero if two punctures are connected by the exact WKB curves. Then, if there exists a contractible cycle connecting the same puncture $z_i$, the Wronskian between two identical small solutions $\sl{s_i\comma s_i}$ must be nonzero. However, this obviously contradicts the definition of the Wronskian. On the other hand, if the curve connecting the same puncture is non-contractible, namely if the curve goes around a different puncture at least once, we need to take into account the effect of the monodromy $\tilde{\Omega}$ around the punctures and the Wronskian which should be nonzero is now replaced by $\sl{s_i \comma \tilde{\Omega} s_i}$. This does not lead to any contradiction and thus we conclude the curves which connect the same puncture should always be non-contractible. The next property to show is given as follows:
\begin{itemize} 
\item[]{\it Endpoints}. All but finitely many exact WKB curves terminate at punctures.
\end{itemize}
To derive this, let us first classify possible endpoints of the exact WKB curves. As in the case of ordinary WKB curves, possible endpoints are zeros or poles of $P_{\rm ex}^{(i)}$. Among these two, zeros are unimportant since the number of the exact WKB curves flowing into zeros is always finite as shown in \figref{fig-WKB}. On the other hand, a pole can be an endpoint for infinitely many curves and thus plays a crucial role in the analysis. There are basically three different types for poles. The first type is the punctures, at which the vertex operators are inserted.
The second type is the singularities of the small eigenvector $s_i$, different from the punctures. However, since the auxiliary linear equation we are solving is of the Fuchsian type, we expect $s_i$ becomes singular only at the punctures and other singularities are absent. The last type is the zeros of the eigenvector $s_i$. Although $s_i$ in general has several zeros on the Riemann surface, such points cannot be the endpoints of the exact WKB curves for the following reason: At zeros of $s_i$, the ratio between $s_i$ and $b_i$, $s_i/b_i$, must also be zero\fn{$b_i$ must be nonzero at such points to ensure the normalization condition $\sl{s_i\comma b_i}=1$.}. However, this contradicts the basic property of the exact WKB curve that such a ratio, determined by the exponential factor in \eqref{si}, monotonically increases along the exact WKB curve as we move away from $z_i$.
  From these considerations, we can conclude that all but finitely many exact WKB curves end at punctures.

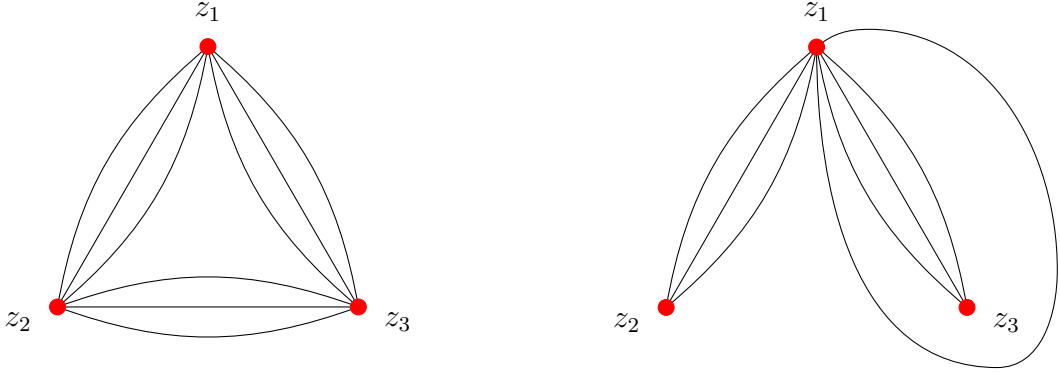
\begin{figure}
\begin{minipage}{0.5\hsize}
\centering
\begin{tikzpicture}
\draw (0,0) -- (2,{2*sqrt(3)});
\draw (0,0) to [out=80, in=220] (2,{2*sqrt(3)});
\draw (0,0) to [out=40, in=260]  (2,{2*sqrt(3)});
\draw (4,0) -- (2,{2*sqrt(3)});
\draw (4,0) to [out=100, in=320] (2,{2*sqrt(3)});
\draw (4,0) to [out=140, in=280] (2,{2*sqrt(3)});
\draw (0,0) -- (4,0);
\draw (0,0) to [out=20, in=160] (4,0);
\draw (0,0) to [out=340, in=200] (4,0);
\filldraw[red] (0,0) circle (3pt)
(2,{2*sqrt(3)}) circle (3pt)
(4,0) circle (3pt);
\node[below,left] at (-0.2,-0.2) {$z_2$};
\node[above] at (2, {2*sqrt(3)+0.2}) {$z_1$};
\node[below,right] at (4.2,-0.2) {$z_3$};
\node[below] at (2,-1) {(a) Symmetric case ($N_i + N_j-N_k>0$)};
\end{tikzpicture}
\end{minipage}
\begin{minipage}{0.5\hsize}
\centering
\begin{tikzpicture}
\draw (0,0) -- (2,{2*sqrt(3)});
\draw (0,0) to [out=80, in=220] (2,{2*sqrt(3)});
\draw (0,0) to [out=40, in=260]  (2,{2*sqrt(3)});
\draw (4,0) -- (2,{2*sqrt(3)});
\draw (4,0) to [out=100, in=320] (2,{2*sqrt(3)});
\draw (4,0) to [out=140, in=280] (2,{2*sqrt(3)});
\draw (2,{2*sqrt(3)}) to [out=270, in=180] (4.4,-0.8) to [out=0, in=270] (5.2,0.5) to [out=90, in=0] (2.7,3.7) to [out=180, in=40] (2,{2*sqrt(3)});
\filldraw[red] (0,0) circle (3pt)
(2,{2*sqrt(3)}) circle (3pt)
(4,0) circle (3pt);
\node[below,left] at (-0.2,-0.2) {$z_2$};
\node[above] at (2, {2*sqrt(3)+0.2}) {$z_1$};
\node[below,right] at (4.2,-0.2) {$z_3$};
\node[below] at (2,-1) {(b) Asymmetric case ($N_2 + N_3-N_1<0$)};
\end{tikzpicture}
\end{minipage}
\caption{Two types of configurations of the exact WKB curves.
In the symmetric case, all the punctures are connected to each other by exact WKB curves.
In the asymmetric case, we have non-contractible exact WKB curves which start and end at the same puncture. As the exact WKB curves are non-intersecting, there exists a pair of punctures between which there are no connecting curves in such a case.}
\label{fig-connection}
\end{figure}

Having proven two important properties, 
now we are in a position to discuss the configuration of the exact WKB curves. There are essentially two different cases. First consider the case where $N_i$'s satisfy the triangle inequality:
\beq{
N_i + N_j-N_k>0\period
}
In this case, to be called the {\it symmetric case}, the only allowed configuration which obeys the properties derived above is the one in which three punctures are piece-wise connected (see \figref{fig-connection}). Next consider the case where the following triangular inequality is violated:
\beq{
N_2+N_3-N_1<0\period \label{TIv}
}
Note that all the remaining cases can be obtained from \eqref{TIv} by the permutation of three indices $1,2$ and $3$. These cases are called the {\it asymmetric cases}. In the case \eqref{TIv}, the only consistent way to connect three punctures is to arrange the curves so that all the curves emanating from $z_2$ and $z_3$ end at $z_1$ and some of the curves emanating from $z_1$ go back to $z_1$ by going around $z_2$ or $z_3$ (see \figref{fig-connection}).

In this way, we can completely determine the configuration of the exact WKB curves by the local data of the auxiliary linear problem. Below, we will see explicitly how the configuration of the exact WKB curves can be used to determine zeros of the Wronskians.
\subsubsection{Determination of zeros}
To illustrate the outline of the discussion with an explicit example, let us consider the factor,
\beq{
\sin \left( \frac{p_1 +p_2 +p_3}{2}\right) \comma\label{plplpl}
}
and determine which of the Wronskian become zero when \eqref{plplpl} vanishes.
From \eqref{WR-Product1}--\eqref{WR-Product6}, the products of the Wronskians which vanish are given by
\beq{
\sl{1_{+}\comma 2_{+}}\sl{1_{-}\comma 2_{-}}\comma \quad \sl{2_{+}\comma 3_{+}}\sl{2_{-}\comma 3_{-}}\comma \quad\sl{3_{+}\comma 1_{+}}\sl{3_{-}\comma 1_{-}}\period\label{group}
}
An important feature of \eqref{group} is that, if we group the eigenvectors into $\{1_{+},2_{+},3_{+}\}$ and $\{1_{-},2_{-},3_{-}\}$, only the Wronskians within the same group appear in \eqref{group}. This is in fact quite a general feature and is true also for other situations. 

Now, let us prove two theorems, which will be useful for determination of zeros. The first theorem is the following one, which we have already proven:
\begin{itemize}
\item[]{\it Theorem 1.}
When two punctures $z_i$ and $z_j$ are connected by the exact WKB curve, the Wronskian between two small eigenvectors $\sl{s_i \comma s_j}$ is nonzero.
\end{itemize}
Before writing down the second theorem, let us prove the following useful lemma:
\begin{itemize}
\item[]{\it Lemma.}
In \eqref{group}, only one of the two Wronskians in each product vanishes. The other does not vanish.
\end{itemize}
To prove {\it Lemma}, first note that all the zeros of \eqref{plplpl} are simple zeros. Note also that the eigenvectors and their Wronskians are in general single-valued around the zeros of \eqref{plplpl}. From these two facts, it follows that only one of the two Wronskians in the product vanishes at the zeros of \eqref{plplpl}. If both of the Wronskians vanish, \eqref{plplpl} must have a double zero and contradicts the aforementioned property. Now, using these lemmas, we can derive the following important theorem.
\begin{itemize}
\item[]{\it Theorem 2.}
There are only two distinct possibilities concerning zeros of the Wronskians in \eqref{group}:\\
1.~All the Wronskians among $\{1_{+},2_{+},3_{+}\}$ are zero and all the Wronskians among $\{1_{-},2_{-},3_{-}\}$ are nonzero.\\
2.~All the Wronskians among $\{1_{+},2_{+},3_{+}\}$ are nonzero and all the Wronskians among $\{1_{-},2_{-},3_{-}\}$ are zero.
\end{itemize}
This theorem can be proven in the following way. First, assume that two of the three Wronskians in $\{1_{+},2_{+},3_{+}\}$ are zero. In such a case,  
all $i_{+}$'s  become proportional to each other. Then, it follows that the remaining one Wronskian also vanishes. Using this line of reasoning and the results of {\it Lemma}, we can easily show that there are only two distinct possibilities, described in {\it Theorem 2}.

Using the theorems we just proved, we can determine zeros of the Wronskians at relative ease when the configuration of the exact WKB curve is symmetric. In such cases, one of the groups, $\{1_{+},2_{+},3_{+}\}$ and $\{1_{-},2_{-},3_{-}\}$, contains at least two small eigenvectors. Then, applying {\it Theorem 1} and {\it Theorem 2}, we can immediately conclude that the Wronskians from such a group are all nonzero whereas the Wronskians from the other group are all zero. Let us next consider the asymmetric case. For simplicity, let us assume that $N_1>N_2 + N_3$ is satisfied\fn{Generalization to other cases is straightforward.}. In such a case, there exist exact WKB curves which start $z_1$, go around $z_2$ (or $z_3$) and return to $z_1$. Therefore, to utilize the information on the exact WKB curves, it is important to consider the Wronskians,
\beq{
\sl{1_+ \comma \Omega_2 1_+} \comma \quad \sl{1_{-}\comma \Omega_2 1_{-}}\period\label{asymwron}
}
To compute these Wronskians, we first note $1_\pm$ can be expressed in terms of $2_{\pm}$ as follows:
\beq{
1_{\pm} = \sl{1_{\pm}\comma 2_{-}} 2_{+}-\sl{1_{\pm}\comma 2_{+}} 2_{-}\period\label{1intermsof2}
}
Then, applying $\Omega_2$ to \eqref{1intermsof2} and substituting them to \eqref{asymwron}, we can express \eqref{asymwron} in terms of the ordinary Wronskians as
\beq{
&\sl{1_+ \comma \Omega_2 1_+} = 2i \sin p_2 \sl{1_{+}\comma 2_{-}}\sl{1_{+}\comma 2_{+}}\comma\label{1om1}\\
&\sl{1_- \comma \Omega_2 1_-}= 2i \sin p_2 \sl{1_{-}\comma 2_{-}}\sl{1_{-}\comma 2_{+}}\period
}
Since $\Omega_2 1_{+}$ can be obtained by parallel-transporting $1_{+}$ along the exact WKB curve which starts and ends at $z_1$, $\Omega_2 1_{+}$ behaves as the big solution around $z_1$ when $1_{+}$ is the small solution. Therefore, the Wronskian $\sl{1_+ \comma \Omega_2 1_+} $ is nonzero in this case. Then, because of the expression \eqref{1om1}, $\sl{1_{+}\comma 2_{+}}$ is also nonzero. Consequently, using {\it Theorem 2}, we can show that the Wronskians among $\{1_{+},2_{+},3_{+}\}$ are nonzero and the Wronskians among $\{1_{-},2_{-},3_{-}\}$ are zero. Similarly, we can show that the Wronskians among $\{1_{-},2_{-},3_{-}\}$ are nonzero and the Wronskians among $\{1_{+},2_{+},3_{+}\}$ are zero when $1_{-}$ is the small eigenvector.

Performing a similar analysis also for other zeros, we can derive the following general rules:
\begin{itemize}
\item[]1. {\it Decomposition of the eigenvectors into two groups.}\\When a factor of the form $\sin \left( \sum_i \epsilon_i p_i/2\right)$ vanishes, the Wronskians which vanish are the ones among $\{1_{\epsilon_1}, 2_{\epsilon_2},3_{\epsilon_3} \}$ and the ones among $\{1_{-\epsilon_1}, 2_{-\epsilon_2},3_{-\epsilon_3} \}$.
\item[]2. {\it Symmetric case.}\\When the configuration of the exact WKB curves is symmetric, the Wronskians from the group which contains two or more small solutions are nonzero whereas the Wronskians from the other group are zero.
\item[]3. {\it Asymmetric case.}\\When the configuration of the exact WKB curves is asymmetric and $N_i$'s satisfy $N_i>N_j+N_k$, the Wronskians from the group which contains the smaller solution of $i_{\pm}$ are nonzero whereas the Wronskians from the other group are zero.
\end{itemize}

Let us now examine the above rules more in detail. For simplicity, let us focus on $\sl{1_{+}\comma 2_{+}}$ and consider the zeros which arise when $\sin \left(\sum_i p_i /2\right)$ vanishes. To apply the general rules, we need to know 1) Which of $i_{\pm}$ is small and 2) The configuration of the exact WKB curves. What is important in the following discussion is that both of these are determined by the real part of $\eta_i(\xi) -1/2$. In fact, the relative magnitude of $i_{\pm}$ is determined as
\beq{
\begin{aligned}
&\Re \left(\eta_i(\xi) -\frac{1}{2} \right)>0\quad\Rightarrow\quad i_{+}: \text{small}\comma \,i_{-}: \text{big}\comma\\
&\Re \left(\eta_i(\xi) -\frac{1}{2} \right)<0\quad\Rightarrow\quad i_{+}: \text{big}\comma \,i_{-}: \text{small}\comma 
\end{aligned}
}
and the numbers of the exact WKB curves $N_i$ are given in terms of $\eta_i(\xi) -1/2$ as shown in \eqref{defni}. Below we will denote this important quantity by $\gamma_i$, $\gamma_i\equiv \Re\left( \eta_i(\xi) -1/2\right)$. Let us first consider the case where all $\gamma_i$'s are positive. In this case, $1_{+}$, $2_{+}$ and $3_{+}$ become all small and $\sl{1_{+}\comma 2_{+}}$ is nonzero irrespective of the configuration of the exact WKB curves. Next consider the case where only $\gamma_3$ is negative and $3_+$ is a big solution. In this case, $\sl{1_{+}\comma 2_{+}}$ vanishes if and only if $N_3>N_1+N_2$ is satisfied. This condition translates into the following condition for $\gamma_i$: $\gamma_1 + \gamma_2 + \gamma_3 <0$. Similarly, in the case where only $\gamma_1$ (or $\gamma_2$) is negative, $\sl{1_{+}\comma 2_{+}}$ vanishes when $N_1>N_2 +N_3$ (or $N_2>N_3 +N_1$) is satisfied. This condition also translates into the same condition, $\gamma_1 + \gamma_2 + \gamma_3 <0
 $.
  One can perform this kind of analysis also for the cases where two or more $\gamma_i$'s are negative and, as a result, we find that zeros of $\sl{1_{+}\comma 2_{+}}$ associated with $\sin \left( \sum_i p_i /2\right)=0$ are completely determined by the sign of $\gamma_1 + \gamma_2 + \gamma_3 $. Re-expressing this condition in terms of $p_i$'s, we obtain the following rule
\beq{
\begin{aligned}
&\Re \left( \frac{p_1 +p_2 +p_3}{2}\right)>\frac{3\pi}{2}\quad \Rightarrow \quad \sl{1_{+}\comma 2_{+}} \neq 0\comma\\
& \Re \left(  \frac{p_1 +p_2 +p_3}{2}\right)<\frac{3\pi}{2}\quad \Rightarrow \quad \sl{1_{+}\comma 2_{+}} = 0\period
\end{aligned}
}
Applying this logic to other situations, we can get explicit rules to determine zeros of the Wronskians.

Now, let us write down the final result combining the results for poles and the results for zeros. For simplicity, here we just present the result for $\langle 1_{\pm }\comma 2_{\pm}\rangle$, but the analyticity of other Wronskians can be easily obtained by the permutation of indices, 1, 2 and 3.
\beq{
&\text{{\bfall Analyticity of $\langle 1_{+}\comma 2_{+}\rangle$}}\nn\\
&\hspace{10pt}\text{Poles: At $\sin p_1 =0$ if $\Re \, p_1<\pi$, and at $\sin p_2 =0$ if $\Re \, p_2<\pi$}\comma\nn\\
&\hspace{10pt}\text{Zeros: At $\sin \frac{p_1+p_2-p_3}{2}=0$ if $\Re \, \frac{p_1+p_2-p_3}{2} <\frac{\pi}{2}$}\comma\\
&\hspace{10pt}\text{\hspace{40pt}and at $\sin \frac{p_1+p_2+p_3}{2}=0$ if $\Re \, \frac{p_1+p_2+p_3}{2} <\frac{3\pi}{2}$}\period\nn\\
&\text{{\bfall Analyticity of $\langle 1_{-}\comma 2_{-}\rangle$}}\nn\\
&\hspace{10pt}\text{Poles: At $\sin p_1 =0$ if $\Re \, p_1\geq\pi$, and at $\sin p_2 =0$ if $\Re \, p_2\geq\pi$}\comma\nn\\
&\hspace{10pt}\text{Zeros: At $\sin \frac{p_1+p_2-p_3}{2}=0$ if $\Re \, \frac{p_1+p_2-p_3}{2} >\frac{\pi}{2}$}\comma\\
&\hspace{10pt}\text{\hspace{40pt}and at $\sin \frac{p_1+p_2+p_3}{2}=0$ if $\Re \, \frac{p_1+p_2+p_3}{2} >\frac{3\pi}{2}$}\period\nn\\
&\text{{\bfall Analyticity of $\langle 1_{+}\comma 2_{-}\rangle$}}\nn\\
&\hspace{10pt}\text{Poles: At $\sin p_1 =0$ if $\Re \, p_1<\pi$, and at $\sin p_2 =0$ if $\Re \, p_2\geq\pi$}\comma\nn\\
&\hspace{10pt}\text{Zeros: At $\sin \frac{-p_1+p_2+p_3}{2}=0$ if $\Re \, \frac{-p_1+p_2+p_3}{2} >\frac{\pi}{2}$}\comma\\
&\hspace{10pt}\text{\hspace{40pt}and at $\sin \frac{p_1-p_2+p_3}{2}=0$ if $\Re \, \frac{p_1-p_2+p_3}{2} <\frac{\pi}{2}$}\period\nn\\
&\text{{\bfall Analyticity of $\langle 1_{-}\comma 2_{+}\rangle$}}\nn\\
&\hspace{10pt}\text{Poles: At $\sin p_1 =0$ if $\Re \, p_1\geq\pi$, and at $\sin p_2 =0$ if $\Re \, p_2<\pi$}\comma\nn\\
&\hspace{10pt}\text{Zeros: At $\sin \frac{-p_1+p_2+p_3}{2}=0$ if $\Re \, \frac{-p_1+p_2+p_3}{2} <\frac{\pi}{2}$}\comma\\
&\hspace{10pt}\text{\hspace{40pt}and at $\sin \frac{p_1-p_2+p_3}{2}=0$ if $\Re \, \frac{p_1-p_2+p_3}{2} >\frac{\pi}{2}$}\period\nn
}
In the next subsection, we will utilize this information to explicitly compute the individual Wronskians.
\subsection{Calculation of Wronskians and three-point functions}
\begin{figure}
\centering
\begin{tikzpicture}
\draw[thick,loosely dashed] (0,-1) -- (0,1);
\draw[thick,
    decoration={markings,mark=at position 1 with {\arrow[scale=2]{>}}},
    postaction={decorate},
    shorten >=0.4pt
    ] (-5,0) -- (5,0);
\foreach \x in {0,...,4}
{\filldraw[red] ({0.5+\x},0) circle (3pt);
\filldraw[blue] ({-0.5-\x},0) circle (3pt);}
\node[below] at (0,-1) {$z=\frac{\pi}{2}$};
\node[above] at (3,0.5) {Zeros of $1/\Gamma(1-\frac{z}{\pi})$};
\node[above] at (-3,0.5) {Zeros of $1/\Gamma(\frac{z}{\pi})$};
\end{tikzpicture}
\caption{Decomposition of the zeros of $\sin z$. $1/\Gamma (1-\frac{z}{\pi})$ contains zeros on the positive real axis whereas $1/\Gamma(\frac{z}{\pi})$ contains zeros on the negative real axis.}
\label{fig-zeros}
\end{figure}
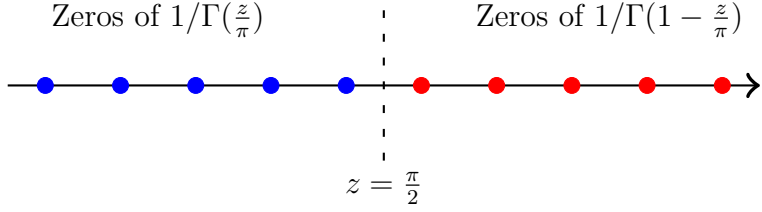

Now let us evaluate Wronskians. To evaluate Wronskians, we need to decompose the right hand sides of \eqref{WR-Product1}-\eqref{WR-Product6} based on the analyticity properties listed above. To carry this out, we will make use of the following useful formula:
\beq{
\sin z = \frac{\pi}{\Gamma (\frac{z}{\pi})\Gamma( 1-\frac{z}{\pi})}\period \label{WR-GammaFormula}
}
Since the gamma function has poles only at $n\in \mathbb{Z}_{\leq 0}$, \eqref{WR-GammaFormula} decomposes the zeros of $\sin z$ into two groups, those on the positive real axis and those on the negative real axis (see \figref{fig-zeros}). 
Then, from the analyticity conditions determined in the previous subsection, we obtain the following decompositions:
\beq{
&\sl{1_{+}\comma 2_{+}} \propto \frac{\Gamma (p_1/\pi )\Gamma (p_2/\pi )}{\Gamma \left( \frac{p_1+p_2-p_3}{2\pi}\right)\Gamma \left( \frac{p_1+p_2+p_3}{2\pi}-1\right)}\comma\label{WR-1P2P}\\
&\sl{1_{-}\comma 2_{-}} \propto \frac{\Gamma (1- p_1/\pi )\Gamma (1- p_2/\pi )}{\Gamma \left( 1- \frac{p_1+p_2-p_3}{2\pi}\right)\Gamma \left( 2-\frac{p_1+p_2+p_3}{2\pi}\right)}\comma\label{WR-1M2M}\\
&\sl{1_{+}\comma 2_{-}} \propto \frac{\Gamma (p_1/\pi )\Gamma (1-p_2/\pi )}{\Gamma \left( 1-\frac{-p_1+p_2+p_3}{2\pi}\right)\Gamma \left( \frac{p_1-p_2+p_3}{2\pi}\right)}\comma\\
&\sl{1_{-}\comma 2_{+}} \propto \frac{\Gamma (1-p_1/\pi )\Gamma (p_2/\pi )}{\Gamma \left( \frac{-p_1+p_2+p_3}{2\pi}\right)\Gamma \left( 1-\frac{p_1-p_2+p_3}{2\pi}\right)}\period
}
Note that the constants of proportionality in the above equations cannot be determined purely by the analyticity.

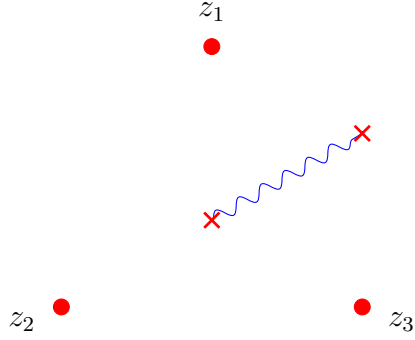
\begin{figure}
\centering
\begin{tikzpicture}
\filldraw[red] (0,0) circle (3pt)
(2,{2*sqrt(3)}) circle (3pt)
(4,0) circle (3pt);
\draw[snake=snake, blue]  (2,{2*sqrt(3)/3}) -- (4,{4*sqrt(3)/3});
\node[below,left] at (-0.2,-0.2) {$z_2$};
\node[above] at (2, {2*sqrt(3)+0.2}) {$z_1$};
\node[below,right] at (4.2,-0.2) {$z_3$};
\node at (2,{2*sqrt(3)/3}) {\Cross};
\node at (4,{4*sqrt(3)/3}) {\Cross};
\end{tikzpicture}
\caption{Branch cut of $\sqrt{-T(z)}$. We choose $\eta_i$ such that the branch cut is located between $z_1$ and $z_2$.}
\label{fig-branch}
\end{figure}

To determine the proportionality constants, we study the asymptotic behavior of Wronskians as $\xi$ goes to infinity using the WKB expansion. As the relative magnitude of $i_{\pm}$ depends on $\text{Arg }\xi$, we choose $\text{Arg }\xi$ appropriately such that $i_{+}$ is the small solution and $i_{-}$ is the big solution in the limit $\xi \to \infty$. In such a case, $i_{+}$ can be expressed in terms of $P_{\rm ex}^{(i)}$ as \eqref{si}. To determine the constant of proportionality in \eqref{si}, we study the behavior around $z_i$ and use the approximate form\fn{In this case, the minus sign must be chosen in \eqref{appex} to reproduce the behavior of $i_{+}$ near $z_i$.} of $P_{\rm ex}^{(i)}$ in the vicinity of $z_i$, \eqref{appex}. Then, the constant of proportionality can be computed by comparing the definition of the right hand side of \eqref{si} and the normalization of $i_{+}$, given in \eqref{WR-HolAsymptMod}. The result is given as follows:
\beq{
i_{+} &= \lim_{\epsilon_i \to 0} \epsilon_i^{\eta_i(\xi)-1/2} \sqrt{\frac{1/2-\eta_i(\xi )}{P_{{\rm ex}}^{(i)} (z,\xi)}}\exp \left( - \int ^{z}_{z_i+\epsilon_i} P_{{\rm ex}}^{(i)} (z',\xi)dz^{\prime} \right)\period\label{asympp}
}
Then, approximating $P_{\rm ex}^{(i)}$ in \eqref{asympp} by the leading term of the WKB expansion, we obtain the approximate expression for $i_{+}$ at $\xi\to 0$. However, to write down an explicit expression, we first need to clarify the choice of branch of $\sqrt{-T(z)}$ appearing in the expansion of $P_{\rm ex}^{(i)}$. In the case of the three-point functions, $\sqrt{-T(z)}$ always has one branch cut and its position depends on the relative magnitudes of $\eta_i(1-\eta_i)$. In what follows, we only consider the case where $\eta_2 (1-\eta_2)$ is larger than $\eta_1 (1-\eta_1)$ and $\eta_3 (1-\eta_3)$ and the branch cut is located between $z_1$ and $z_3$, as shown in \figref{fig-branch}. In such a case, the branch of $\sqrt{-T(z)}$ on the first sheet can be chosen as follows:
\beq{
\sqrt{-T(z)}&\sim \frac{i\left|\sqrt{\eta_i(1-\eta_i)}\right|}{z-z_i}\quad \text{as }z\to z_i\quad \text{for }i=1,3\comma\\
&\sim \frac{-i\left|\sqrt{\eta_i(1-\eta_i)}\right|}{z-z_2}\quad \text{as }z\to z_2 \period
}
Although the discussion below is restricted to this particular case, the final result will turn out to be completely symmetric under the permutation of $z_1$, $z_2$ and $z_3$ and is applicable also to other cases.

With this choice of the branch, we can write down approximate expressions for $i_{+}$ as
\beq{
&1_{+}\sim  \epsilon_1^{\eta_1 (\xi)-1/2}\sqrt{\frac{1}{2}-\eta_1(\xi)} \left( \frac{\xi^2}{-T}\right)^{1/4}\exp \left(-\frac{1}{\xi}\int_{z_1 + \epsilon _1}^{z} \sqrt{-T}dz^{\prime} \right)\comma\\
&2_{+}\sim \epsilon_2^{\eta_2 (\xi)-1/2}\sqrt{\frac{1}{2}-\eta_2(\xi)} \left( \frac{\xi^2}{-T}\right)^{1/4}\exp \left(\frac{1}{\xi}\int_{z_2 +\epsilon_2}^{z} \sqrt{-T}dz^{\prime} \right)\comma\\
&3_{+}\sim \epsilon_3^{\eta_3 (\xi)-1/2}\sqrt{\frac{1}{2}-\eta_3(\xi)} \left( \frac{\xi^2}{-T}\right)^{1/4}\exp \left(-\frac{1}{\xi}\int_{z_3 + \epsilon _3}^{z} \sqrt{-T}dz^{\prime} \right)\period
}
Then the Wronskians which only include $i_{+}$ can be calculated as
\beq{
\sl{1_{+}\comma 2_{+}}\sim & \lim_{\epsilon_i \to 0} \epsilon_1^{\eta_1(\xi)-1/2} \epsilon_2^ {\eta_2(\xi)-1/2}\sqrt{\left( 1-2\eta_1(\xi)\right)\left( 1-2\eta_2(\xi)\right)} \exp \left( - \frac{1}{\xi}\int _{z_1 +\epsilon_1}^{z_2 + \epsilon _2} \sqrt{-T}dz^{\prime} \right)\comma\\
\sl{2_{+}\comma 3_{+}}\sim & \lim_{\epsilon_i \to 0} \epsilon_2^{\eta_2(\xi)-1/2}\epsilon_3^{\eta_3(\xi)-1/2} \sqrt{\left( 1-2\eta_2(\xi)\right)\left( 1-2\eta_3(\xi)\right)} 
\exp \left( - \frac{1}{\xi}\int _{z_3 +\epsilon_2}^{z_2 + \epsilon _3} \sqrt{-T}dz^{\prime} \right)\comma\\
\sl{3_{+}\comma 1_{+}}\sim & \lim_{\epsilon_i \to 0} \epsilon_3^{\eta_3(\xi)-1/2} \epsilon_1^{\eta_1(\xi)-1/2} \sqrt{\left( 1-2\eta_3(\xi)\right)\left( 1-2\eta_1(\xi)\right)}\exp \left( - \frac{1}{\xi}\int _{z_3 +\epsilon_3}^{\hat{z}_1 + \epsilon _1} \sqrt{-T}dz^{\prime} \right)\comma
}
where $\hat{z}_1$ in the last line denotes the point on the lower sheet of $y^2 =-T(z)$ right below $z_1$.
To evaluate the above formulae, we need to compute the contour integral $\int _{z_i}^{z_j} \sqrt{-T} dz^{\prime}$.
The easiest way to compute them is to map $z_{1}$, $z_{2}$ and $z_3$ to $0$, $1$ and $\infty$ by the ${\rm SL}(2,\mathbb{C})$-transformation and then evaluate the integral taking into account the effect of the transformation. The transformation we use is
\beq{
w \equiv -\frac{z_{23}}{z_{12}} \frac{z-z_1}{z-z_3}\comma
}
where $z_{ij}\equiv z_i-z_j$. The stress-energy tensor in this coordinate is given by
\beq{
T(w) = \frac{\eta_1 (1-\eta_1 )+ \left(\eta_2 (1-\eta_2 )-\eta_1 (1-\eta_1 )-\eta_3 (1-\eta_3 ) \right)w +\eta_3 (1-\eta_3 ) w^2}{w^2 (w-1)^2}\period
}
Then, the Wronskians can be re-expressed as
\beq{
\sl{1_{+}\comma 2_{+}}\sim & \left(\frac{z_{12} z_{31}}{z_{23}}\right)^{\eta_1 (\xi )-1/2}\left(\frac{z_{12}z_{23}}{z_{31}}\right)^{\eta_2 (\xi )-1/2}\sqrt{\left( 1-2\eta_1(\xi)\right)\left( 1 - 2\eta_2(\xi)\right)}\nn\\
&\times \lim_{\epsilon_i \to 0}  \epsilon_1^{\eta_1(\xi)-1/2} \epsilon_2^ {\eta_2(\xi)-1/2} \exp \left( -\frac{1}{\xi}\int _{\epsilon_1}^{1+ \epsilon _2} \sqrt{-T}dw^{\prime} \right)\comma\\
\sl{2_{+}\comma 3_{+}}\sim & \left(\frac{z_{12} z_{23}}{z_{31}}\right)^{\eta_2 (\xi )-1/2}\left(\frac{z_{23}z_{31}}{z_{12}}\right)^{\eta_3 (\xi )-1/2} \sqrt{\left( 1- 2\eta_2(\xi)\right)\left( 1-2\eta_3(\xi)\right)}\nn\\
& \lim_{\epsilon_i \to 0} \epsilon_2^{\eta_2(\xi)-1/2} \epsilon_3^{\eta_3(\xi)-1/2} \exp \left( -\frac{1}{\xi}\int _{-\epsilon_3^{-1}}^{1 + \epsilon _2} \sqrt{-T}dw^{\prime} \right)\comma\\
\sl{3_{+}\comma 1_{+}}\sim & \left(\frac{z_{23}z_{31}}{z_{12}}\right)^{\eta_3 (\xi )-1/2}\left(\frac{z_{12} z_{31}}{z_{23}}\right)^{\eta_1 (\xi )-1/2}\sqrt{\left( 1 - 2\eta_3(\xi)\right)\left( 1 - 2\eta_1(\xi)\right)} \nn\\
&\lim_{\epsilon_i \to 0} \epsilon_3^{\eta_3(\xi)-1/2} \epsilon_1^{\eta_1(\xi)-1/2} \exp \left( -\frac{1}{\xi}\int _{-\epsilon_3^{-1}}^{\hat{0}+ \epsilon _1} \sqrt{-T}dw^{\prime} \right)\comma
}
where the regularization parameters $\epsilon_i$'s in the above formulae are redefined as 
\beq{
&\epsilon_1 ^{\rm new}\equiv \frac{z_{23}}{z_{12}z_{31}}\epsilon_1^{\rm old}\comma\quad
\epsilon_2 ^{\rm new}\equiv \frac{z_{31}}{z_{12}z_{23}}\epsilon_2^{\rm old}\comma\quad
\epsilon_3 ^{\rm new}\equiv \frac{z_{12}}{z_{23}z_{31}}\epsilon_3^{\rm old}\comma
}
and the prefactors depending on $z_{ij}$ are results of such redefinition.
Each contour integral $\int \sqrt{-T}dw$ is divergent in the limit $\epsilon_i\to 0$. However, combined with the prefactors $\epsilon_{\ast}$, they become completely finite and can be evaluated analytically.
Then, by comparing\fn{To compare two expressions, we used the Stirling formula, $\log\Gamma (x) \sim (x-1/2) \log x -x$.} the resultant asymptotic formula with the $\xi\to 0$ limit of the left hand side of \eqref{WR-1P2P} and \eqref{WR-1M2M}, we can finally determine $\sl{1_{+}\comma 2_{+}}$ and $\sl{1_{-}\comma 2_{-}}$ as
\beq{
&\sl{1_{+}\comma 2_{+}} =\left(\frac{z_{12} z_{31}}{z_{23}}\right)^{p_1/2\pi -1/2}\left(\frac{z_{12} z_{23}}{z_{31}}\right)^{p_2/2\pi-1/2}\frac{\Gamma (p_1/\pi )\Gamma (p_2/\pi )}{\Gamma \left( \frac{p_1+p_2-p_3}{2\pi}\right)\Gamma \left( \frac{p_1+p_2+p_3}{2\pi}-1\right)}\comma\\
&\sl{1_{-}\comma 2_{-}} =\left(\frac{z_{12} z_{31}}{z_{23}}\right)^{1/2-p_1/2\pi}\left(\frac{z_{12} z_{23}}{z_{31}}\right)^{1/2-p_2/2\pi}\frac{\Gamma (1- p_1/\pi )\Gamma (1- p_2/\pi )}{\Gamma \left( 1- \frac{p_1+p_2-p_3}{2\pi}\right)\Gamma \left( 2-\frac{p_1+p_2+p_3}{2\pi}\right)}\period
}
Other $\sl{i_{+}\comma j_{+}}$ and $\sl{i_{-}\comma j_{-}}$ are obtained from the permutation of indices. On the other hand, the Wronskians of the type $\sl{i_{+}\comma j_{-}}$ can be obtained by first specifying the explicit form of the monodromy matrices $\Omega_i$ using the result for $\sl{i_{+}\comma j_{+}}$ and $\sl{i_{-}\comma j_{-}}$ and then calculating the Wronskians explicitly. The result is
\beq{
&\sl{1_{+}\comma 2_{-}} =\left(\frac{z_{12} z_{31}}{z_{23}}\right)^{p_1/2\pi-1/2}\left(\frac{z_{12} z_{23}}{z_{31}}\right)^{1/2-p_2/2\pi}\frac{\Gamma (p_1/\pi )\Gamma (1-p_2/\pi )}{\Gamma \left( 1-\frac{-p_1+p_2+p_3}{2\pi}\right)\Gamma \left( \frac{p_1-p_2+p_3}{2\pi}\right)}\comma\\
&\sl{1_{-}\comma 2_{+}} =\left(\frac{z_{12} z_{31}}{z_{23}}\right)^{1/2-p_1/2\pi}\left(\frac{z_{12} z_{23}}{z_{31}}\right)^{p_2/2\pi-1/2}\frac{\Gamma (1-p_1/\pi )\Gamma (p_2/\pi )}{\Gamma \left( \frac{-p_1+p_2+p_3}{2\pi}\right)\Gamma \left( 1-\frac{p_1-p_2+p_3}{2\pi}\right)}\period
}
Using these formulae, we can explicitly calculate $C_{i}$ using \eqref{CL-C1} as
\beq{
C_1 &= -\frac{1}{2}\log\lambda -\frac{1-2\eta_1}{2}\log \frac{|z_{12}|^2|z_{31}|^2}{|z_{23}|^2}\cr
&\qquad -\frac{1}{2}\log \frac{\gamma (\eta_1+\eta_2-\eta_3 )\gamma(\eta_3+\eta_1-\eta_2 )\gamma(\eta_1+\eta_2+\eta_3 -1)}{\gamma (2\eta _1)^2 \gamma (\eta_2 + \eta_3 -\eta_1 )}\comma \label{WR-C1} \\
C_2 &= -\frac{1}{2}\log\lambda -\frac{1-2\eta_2}{2}\log \frac{|z_{12}|^2|z_{23}|^2}{|z_{31}|^2}\cr
&\qquad -\frac{1}{2}\log \frac{\gamma (\eta_2+\eta_3-\eta_1 )\gamma(\eta_1+\eta_2-\eta_3 )\gamma(\eta_1+\eta_2+\eta_3 -1)}{\gamma (2\eta _2)^2 \gamma (\eta_3+ \eta_1 -\eta_2 )}\comma \label{WR-C2} \\
C_3 &= -\frac{1}{2}\log\lambda -\frac{1-2\eta_3}{2}\log \frac{|z_{23}|^2|z_{31}|^2}{|z_{12}|^2}\cr
&\qquad -\frac{1}{2}\log \frac{\gamma (\eta_3+\eta_1-\eta_2 )\gamma(\eta_2+\eta_3-\eta_1 )\gamma(\eta_1+\eta_2+\eta_3 -1)}{\gamma (2\eta _3)^2 \gamma (\eta_1 + \eta_2 -\eta_3 )}\comma \label{WR-C3} 
}
where $\gamma(x)$ is defined by
\beq{
\gamma (x)\equiv\frac{\Gamma(x)}{\Gamma (1-x)}\period
}
This result is exactly the same as the one derived by the conventional method in \cite{ZZ,HJ,HMW}. 
Therefore, using the formula \eqref{CL-ActionC}, we can exactly reproduce the classical limit of the DOZZ three-point functions, first derived in \cite{ZZ}. Although the results themselves are already obtained, our method, which does not necessitate the explicit solutions, is quite different from the previous approaches. Accordingly, the exact match of the final results provides strong evidence for the validity and the utility of the method. 

\section{Discussion and future direction}
\label{Sec-Di}
%%%%%%
%In this section we add "Di-" to the head of labels
%%%%%%

In this paper, we have solved the semiclassical three-point functions of the Liouville field theory by utilizing the Riemann-Hilbert analysis.
Instead of directly solving the classical Liouville equation, we have evaluated the Wronskians, which are the bilinears of the solutions to the auxiliary linear problem.
From the monodromy relations around the punctures, we have obtained certain products of the Wronskians.
By applying the method of exact WKB curves \cite{KK3}, we have determined the analyticity of the Wronskians and eventually the exact formula for the Wronskians.
The success in computing the classical Liouville three-point functions strongly supports the validity of this new method.

There are several future directions worth exploring. One is to generalize our method to higher-point functions and conformal blocks. The method used in this paper is closely related to the method used in the holographic calculation of correlation functions in AdS/CFT \cite{JW,KK1,KK2,CT,KK3} and in the wall-crossing phenomena of the $\mathcal{N}=2$ supersymmetric gauge theories \cite{GMN2}.
We expect that TBA-like equations, similar to those obtained in \cite{CT, GMN2}, will be useful to characterize those quantities.
In particular, in the cases involving one irregular puncture \cite{Gai, GT},
we will be able to derive the famous Y-systems as in \cite{AM, AGM, AMSV}. This will also be related to the so-called ODE/IM correspondence\fn{The generalization of the ODE/IM correspondence and its relation to the Liouville field theory is discussed also in \cite{Luk} in a slightly different manner. It would be interesting to clarify the relation between two approaches.}\cite{ODE/IM}, a correspondence between ordinary differential equations and the functional equations in the quantum integrable models such as T-systems and Y-systems. Such investigations are now in progress \cite{WIP}. Another important problem is to study the quantum three-point functions with our method. It would be particularly interesting if we can integrate our method with a novel approach to the quantum three-point functions \cite{CER}, which is based on the non-commutative (quantum) spectral curve.

It is also of interest to explore the relation of our method to the four-dimensional $\mathcal{N}=2$ supersymmetric gauge theories on the omega-deformed background. In the correspondence proposed in \cite{AGT}, 
the parameter $b$ in the Liouville field theory corresponds to one of the parameters of the omega-deformation in the gauge theories. Then the semiclassical limit of the Liouville field theory corresponds to a well-known limit in the gauge theory, known as the Nekrasov-Shatashvili limit, where one of the omega-deformation parameters goes to zero. In this limit, an intriguing relation between the gauge theories and the quantum integrable models has been discovered \cite{NS,MiroMoro,MiroMoroShak,NekWit,Tes2010,MaruTaki,Pogh,MarsMiroMoro,NekRosSha,DLH,CDHL,FMPP,BMT,Bour1,BCGK,FMP,Bour2,CHK,MY, NPS}. This relation, together with the AGT correspondence, implies the existence of the correspondence between the semiclassical Liouville field theory and the quantum integrable models. Although one can derive various nontrivial equalities exploiting such a correspondence \cite{Pia,FerPia}, the origin of the correspondence is still unclear. It would be very interesting if we can get some clues
  for this enigmatic correspondence by investigating the relation between the aforementioned TBA-like equations 
 and
  the quantum integrability in the Nekrasov-Shatashvili limit.

From a mathematical point of view, the Wronskians we have computed are connection coefficients of the hypergeometric differential equations. 
It might be possible to apply our method to more general differential equations and compute their connection coefficients\fn{For example, it would be interesting if the method of this paper can be applied to multivariate hypergeometric functions such as the Gel'fand-Kapranov-Zelevinski systems, which often appear in Mirror symmetry \cite{CK}.}. Such connection coefficients play an important role in the analysis of scattering on a black-hole background \cite{CLMR,CLMR2} and it would be interesting to apply our method to that context. 
Finally, it would be important to clarify the relation of our method to the existing non-perturbative methods for quantum mechanics, such as exact WKB analysis \cite{KT,Voros} and the theory of resurgence \cite{DDP,DP}. 

\section*{Acknowledgement}
We thank D.~Gaiotto, Y.~Kazama, T.~Okuda, A.~Sever and P.~Vieira for useful discussions.
In particular, S.K.\ thanks D.~Gaiotto for kindly sharing his unpublished notes.
The research of D.H.\ and S.K.\ is supported in part by JSPS Research Fellowship for Young Scientists, from the Japan Ministry of Education, Culture, Sports, Science and Technology.
We would like to acknowledge the kind hospitality of Yukawa Institite for Theoretical Physics and the Kavli Institute for the Physics and Mathematics of the Universe, where part of this work was done.
In addition, D.H.\ would like to thank the Simons Center for Geometry and Physics, and S.K.\ would like to thank the Perimeter Institute for Theoretical Physics, for hospitality.

\end{document}